\documentclass[10pt,journal,compsoc]{IEEEtran}
\ifCLASSOPTIONcompsoc
  \usepackage[nocompress]{cite}
\else
  \usepackage{cite}
\fi
\ifCLASSINFOpdf
  \usepackage[pdftex]{graphicx}
  \graphicspath{{figures/}{expr/}}
\else
\fi
\usepackage{url}

\usepackage{paralist}
\let\itemize\compactitem
\let\enditemize\endcompactitem
\let\enumerate\compactenum
\let\endenumerate\endcompactenum

\usepackage{multirow}

\begin{document}
%
\title{Improving the Performance and Endurance of Persistent Memory \\with
Loose-Ordering Consistency
\vspace{-0.05in}}
%
%
%
%

\author{Youyou~Lu,~\IEEEmembership{Member,~IEEE,}
        Jiwu~Shu,~\IEEEmembership{Senior~Member,~IEEE,}
	Long~Sun,
        and~Onur~Mutlu,~\IEEEmembership{Senior~Member,~IEEE}
\IEEEcompsocitemizethanks{\IEEEcompsocthanksitem Youyou Lu, Jiwu Shu and
Long Sun are with the Department of Computer Science and Technology,
Tsinghua University, Beijing, China, 100084.\protect\\
E-mail: \{luyouyou, shujw\}@tsinghua.edu.cn,
sun-l12@mails.tsinghua.edu.cn
\IEEEcompsocthanksitem Onur Mutlu is with the Department of Computer Science
at ETH Zurich and Department of Electrical and Computer Engineering at Carnegie
Mellon University.\protect\\
Email: onur.mutlu@inf.ethz.ch}
}

%
%

\markboth{IEEE Transactions on Parallel and Distributed Systems,~Vol.~15, No.~11, November~2016}%
{Shell \MakeLowercase{\textit{et al.}}: Bare Demo of IEEEtran.cls for Computer Society Journals}
%



\IEEEtitleabstractindextext{%

\vspace{-0.2in}
\begin{abstract}

Persistent memory provides high-performance data persistence at main
memory. Memory writes need to be performed in strict order to satisfy
storage consistency requirements and enable correct recovery from system
crashes. Unfortunately, adhering to such a strict order significantly
degrades system performance and persistent memory endurance. This paper
introduces a new mechanism, Loose-Ordering Consistency (LOC), that
satisfies the ordering requirements at significantly lower performance and
endurance loss. LOC consists of two key techniques. First, \emph{Eager
Commit} eliminates the need to perform a persistent commit record write
within a transaction. We do so by ensuring that we can determine the
status of all committed transactions during recovery by storing
necessary metadata information statically with blocks of data written to
memory. Second, \emph{Speculative Persistence} relaxes the write ordering
between transactions by allowing writes to be speculatively written to
persistent memory. A speculative write is made visible to software only
after its associated transaction commits. To enable this, our mechanism
supports the tracking of committed transaction ID and 
multi-versioning in the CPU cache. Our evaluations show that LOC reduces
the average performance overhead of memory persistence from 66.9\% to 34.9\% and the memory
write traffic overhead from 17.1\% to 3.4\% on a variety of workloads.

\end{abstract}

\vspace{-0.1in}}

\maketitle

\IEEEdisplaynontitleabstractindextext

%
\IEEEpeerreviewmaketitle

\vspace{-0.20in}

\IEEEraisesectionheading{\section{Introduction}\label{sec:intro}\vspace{-0.15in}}

\IEEEPARstart{E}{merging} non-volatile memory~(NVM) technologies, such as Phase Change
Memory~(PCM), Spin-Transfer Torque RAM (STT-RAM) and Resistive
RAM~(RRAM), provide DRAM-like byte-addressable access at DRAM-like
latencies and disk-like data persistence.  Since these technologies
have low idle power, high storage density, and good scalability
properties compared to DRAM~\cite{cacm10pcm, mutlu-memcon13, superfri14research}, they
are regarded as potential alternatives to replace or complement
DRAM as the technology used to build main memory~\cite{isca09pcmlee,
  isca09pcmqureshi, isca09pcmzhou, ieeemicro10pcm, ispass13sttram,
  dac09pdram, iccd12hybrid, yoon2015efficient}.  Perhaps even more importantly, the
non-volatility property of these emerging technologies promises to
enable persistent memory, which can store
data persistently at the main memory level at low
latency~\cite{sosp09bpfs, sc11scmfs, eurosys14pmfs, eurosys16hinfs, asplos11nvheaps,
  asplos11mnemosyne, fast11cdds, weed13pm, usenix17octopus}.

Since memory writes in persistent memory are persistent, they need
to be performed atomically and in correct order to ensure \emph{storage
consistency}. Storage consistency ensures atomicity and durability of
storage systems, so that the system is able to correctly recover from
unexpected system crashes~\cite{tods92aries, acm81shadowpaging,
  osdi08txflash, iccd13lighttx, sosp13mars, tctxssd, linuxexpo98journaling,
  fast13ofss, fast14reconfs}, where volatile data gets
lost (See Section 2.2 for details).
In this work, we borrow the \emph{transaction} concept, which is commonly
used to provide storage consistency in traditional storage
systems~\cite{tods92aries, osdi08txflash, sosp13mars}, to manage persistent memory.
A transaction is atomic: either all of its writes
complete and update persistent memory or none. To accomplish this,
both the old and new versions of the data associated with the location
of a write are kept track of within the transaction. The writes within
and across transactions are {\em persisted (i.e., written to
  persistent memory) in strict program order}, to ensure that correct
recovery is possible in the presence of incomplete transactions. As
such, any persistent memory protocol needs to support both {\em
  transaction atomicity} and {\em strict write ordering to persistent
  memory (i.e., persistence ordering)} in order to satisfy
crash consistency requirements.

Traditionally, disk-based storage systems have employed
transaction-based recovery protocols, such as write-ahead
logging~\cite{tods92aries} or shadow paging~\cite{acm81shadowpaging},
to provide both transaction atomicity and persistence ordering. These
protocols maintain 1) two copies/versions of each data written within
a transaction, and 2) a strict write order to the storage device,
which enables the atomic switch from the old version of data to the
new version upon transaction commit.
Persistent memory is much faster than disk-based storage
and has endurance limitations, i.e., a memory cell wears out after a limited
number of writes~\cite{bookpcm, isca09pcmlee}. Traditional transaction-based recovery
protocols, designed with high-latency disk-based storage
systems in mind, are not suitable for persistent memory due to
their large performance and endurance overhead~\cite{sosp09bpfs,
fast11cdds, cmupdltr11clc, asplos12wholesystempersistence, micro13kiln, isca14memorypersistency}.

Transaction support in persistent memory has two major challenges.
First, the boundary of volatility and persistence in persistent memory
lies between the hardware-controlled CPU cache and the
persistent memory. In contrast, in traditional disk-based storage
systems, the boundary between volatility and persistence lies between
the software-controlled main memory and disk storage. While data
writeback from main memory is managed by the operating system software
in traditional systems, enabling transactional protocols to
effectively control the order of writes to persistent storage, data
writeback from the CPU cache is managed by hardware in persistent memory
systems, making it harder to control the order of writes to persistent
memory at low performance overhead. This is because the CPU cache
behavior is opaque to the system and application software. Therefore,
in order to preserve persistence ordering from the CPU cache to
persistent memory, software needs to explicitly include the relatively
costly cache flush (e.g., \textit{clflush}) and memory fence (e.g.,
\textit{mfence}) instructions (at the end of each transaction) to
force the ordering of cache writebacks~\cite{sc11scmfs, asplos11nvheaps,
asplos11mnemosyne, fast11cdds}.  The average overhead of a clflush and mfence
combined together is reported to be 250ns~\cite{asplos11mnemosyne},
which makes this approach costly, given that persistent memory access
times are expected to be on the order of
tens to hundreds of nanoseconds~\cite{isca09pcmlee,isca09pcmqureshi, ispass13sttram}.
Recently, Intel has introduced two new instructions to reduce flush overheads: 1) \emph{clflushopt}, which provides unordered flush, and 2) \emph{clwb}, which enforces data persistence without invalidating cache lines. However, recent research shows that these instructions still incur high performance overhead, e.g., an average slowdown of 7x for write-intensive persistent memory applications~\cite{micro16delegated}.
In addition, frequent flush operations lead to high memory write
traffic, which accelerates the wear-out process of persistent memory,
thereby hurting its endurance.

Second, existing systems reorder operations, including writes, at
multiple levels, especially in the CPU and the cache hierarchy in
order to maximize system performance. For example, writebacks from the
cache are performed in an order that is usually completely different
from the program-specified order of writes. Similarly, the memory
controller can reorder writes to memory to optimize performance (e.g.,
by optimizing row buffer locality~\cite{isca12sms, micro07stfm}, 
bank-level parallelism~\cite{isca08parbs, micro09membank} and
write-to-read turnaround delays~\cite{lee2010dram,
isca14dbi, micro14firm}).
Enforcing a strict order
of writes to persistent memory to preserve storage consistency
eliminates the reordering across not only writes/writebacks but also
limits reordering possibilities across other operations, thereby
significantly degrading the performance. This is because ensuring a
strict order of writes requires 1) flushing dirty data blocks from
each cache level to memory, 2) writing them back to main memory in the
order specified by the transaction at transaction commit time, and 3)
waiting for the completion of all memory writes within the
transaction before a single write for the next transaction can be
performed. Doing so can greatly degrade system performance, by as high as a factor of 10 
for some memory-intensive workloads we evaluate, as we
demonstrate in Section~\ref{subsec:overall}).

\textbf{Our goal} in this paper is to design new mechanisms that
reduce the performance and endurance overhead caused by strict ordering of writes in
persistent memory. To achieve this, we identify different types of
persistence ordering that degrade performance: intra-transaction
ordering (i.e., strict ordering of writes inside a transaction)
and inter-transaction ordering (i.e., strict ordering
of writes between transactions). We observe that relaxing either of
these types of ordering can be achieved without compromising storage
consistency requirements by changing the persistent memory log
organization and providing hardware support in the CPU cache.
Based on this observation, we develop two complementary
techniques that respectively reduce the performance overhead due to
intra- and inter-transaction (tx) ordering requirements. We
call the resulting mechanism {\em Loose-Ordering Consistency (LOC)}
for persistent memory.

LOC consists of two new techniques. First, a new transaction commit
protocol, {\em Eager Commit}, enables the commit of a transaction
without the use of \emph{commit records}, which traditionally are employed
to record the status of each transaction~\cite{tods92aries, osdi08txflash,
  iccd13lighttx, sosp13optfs, tctxssd}. Doing so removes the need to perform
a persistent commit record write at the end of a transaction and
eliminates the intra-transaction ordering requirement, thereby improving
performance. To achieve this, {\em Eager Commit} organizes the memory
log in a static manner and divides it into groups of blocks (i.e., block groups).
In each group, one metadata block is allocated to keep the transaction
metadata of the other (e.g., seven) data blocks in the block group.
The metadata block is stored along with the data blocks in each block group.
This static log organization enables the system to determine
the status of each transaction by inspecting the metadata information
during recovery, without requiring the use/query of a commit record.
Hence, {\em Eager Commit} eliminates the use of commit records
and removes their associated ordering overhead from the critical path of transaction commit.

Second, {\em Speculative Persistence} relaxes the ordering of writes
between transactions by allowing writes to be speculatively written to
persistent memory. This allows data blocks from multiple transactions
to be written to persistent memory, potentially out of the specified
program order. A speculative write is made visible to software {\em
  only after} its associated transaction commits, and transactions
commit in program order. To enable this, our mechanism requires the
tracking of committed transaction ID and support for multi-versioning
in the CPU cache. Hence, {\em Speculative Persistence} ensures that
storage consistency requirements are met while ordering of persistent
memory writes is relaxed, improving performance and endurance.

\textbf{The major contributions} of this paper are as follows:
\begin{itemize}


\item We identify two types of persistence ordering that lead to
  performance degradation in persistent memory: intra-transaction
  ordering and inter-transaction ordering.

\item We introduce a new transaction commit protocol, \emph{Eager
  Commit}, that eliminates the use of commit records (traditionally
  needed for correct recovery from system crash) and thereby reduces the
  overhead due to intra-transaction persistence ordering.

\item We introduce a new technique, \emph{Speculative Persistence},
  that allows writes from different transactions to speculatively
  update persistent memory in any order while making them visible to
  software only in program order, thereby reducing the overhead of
  inter-transaction persistence ordering.

\item We evaluate our proposals and their combination, \emph{Loose-Ordering
  Consistency (LOC)}, with a variety of workloads ranging
  from basic data structures to graph and database workloads. Results
  show that LOC significantly reduces the average performance overhead
  of persistent memory from 66.9\% to 34.9\% and the memory write
  traffic overhead from 17.1\% to 3.4\%.

\end{itemize}

\section{Background and Motivation}
\label{sec:background}

\subsection{Non-volatile Memory}

Emerging byte-addressable non-volatile memory technologies, also
called storage-class memory technologies, have performance
characteristics close to that of DRAM. For example, one
source~\cite{bookpcm} reports a read latency of 85ns and a write
latency of 100-500ns for Phase Change Memory (PCM). Spin-Transfer
Torque RAM (STT-RAM) has lower latency, e.g., less than 20ns for reads
and writes~\cite{bookpcm}. Their DRAM-comparable performance and
better-than-DRAM technology-scalability, which can enable high memory
capacity at low cost, make these technologies promising alternatives
to DRAM~\cite{isca09pcmlee, isca09pcmqureshi, isca09pcmzhou}. As such,
many recent works examined the use of these technologies as part of
main memory~\cite{isca09pcmlee, isca09pcmqureshi, isca09pcmzhou,
  ieeemicro10pcm, ispass13sttram, dac09pdram, iccd12hybrid, meza2012enabling,
  sosp09bpfs, sc11scmfs, eurosys14pmfs, asplos11nvheaps,
  asplos11mnemosyne, fast11cdds, iccd12meza, weed13pm, ics13memorage, micro14firm,
  micro15thynvm, eurosys16hinfs, inflow16onur, inflow16sttram}, providing disk-like data
persistence at DRAM-like latencies.

Most NVM technologies have the \emph{endurance} problem, i.e., each memory
cell can tolerate only a limited number of writes. The reliability of a memory cell
is weakened when the number of writes grows. When the number of writes
approaches the limit, a memory cell is worn out and can not be used for
storing data~\cite{isca09pcmlee}. The lifetime of a NVM cell is sensitive to the write traffic.
Unfortunately, the endurance problem is serious. For instance, PCM and
STT-RAM can tolerate $10^{6} - 10^{8}$ and $10^{12}$ writes,
respectively~\cite{bookpcm, isca09pcmlee, ispass13sttram}.
Controlling the write traffic is an important issue in persistent memory.

\vspace{-0.15in}
\subsection{Storage Consistency}
\label{sec:background-consistency}

\begin{figure}[htb]
\center
\includegraphics[width=0.82\linewidth]{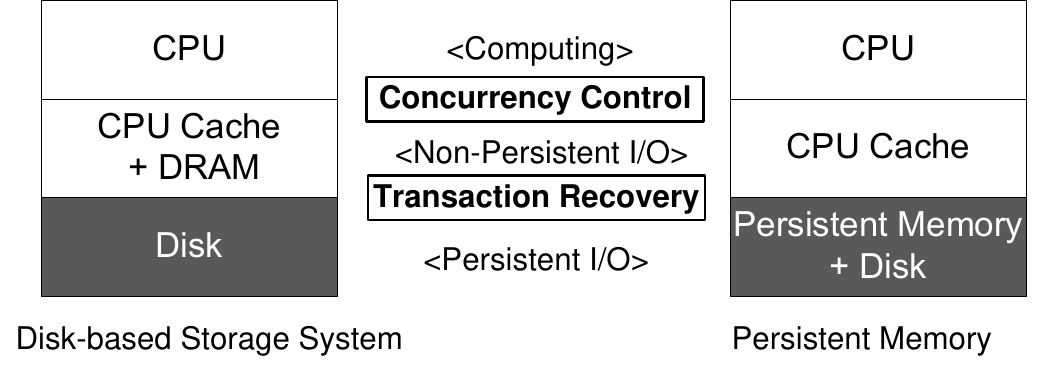}
\vspace{-0.1in}
\caption{Concurrency Control and Transaction Recovery in Disk-based
Storage Systems and Persistent Memory.}
\vspace{-0.1in}
\label{fig:tx-pm}
\end{figure}

\textbf{Storage Transactions.}  In database management
systems, transaction management provides four properties: atomicity
(A), consistency (C), isolation (I), and durability (D). To achieve
the ACID properties, transaction management has to provide 1)
concurrency control for the execution of multiple transactions, 2)
transaction recovery in case of failure. Isolation of concurrent
execution of multiple transactions is the subject of \emph{concurrency
  control}, and the atomic and durable update of state by each
transaction is the subject of \emph{transaction
  recovery}~\cite{book-database}. The two concepts are respectively
borrowed by \emph{transactional memory}~\cite{isca93tm,book:tm2} and
\emph{storage transactions}~\cite{osdi08txflash, iccd13lighttx,
sosp13mars, tctxssd, tcdifftx, linuxexpo98journaling, fast13ofss, nvmsa14txcache}.
Figure~\ref{fig:tx-pm} illustrates the differences of transaction
modules between disk-based storage systems and persistent memory.
While transaction recovery works at the boundary between the DRAM
and disks in disk-based storage systems, it now moves to the boundary between the CPU
cache and the persistent memory in persistent memory systems.
In this paper, we focus on storage transactions (to provide atomicity and
durability guarantees in the presence of system failures) in persistent
memory using the transaction recovery mechanism.

\begin{figure*}
    \centering
    \includegraphics[width=0.79\linewidth]{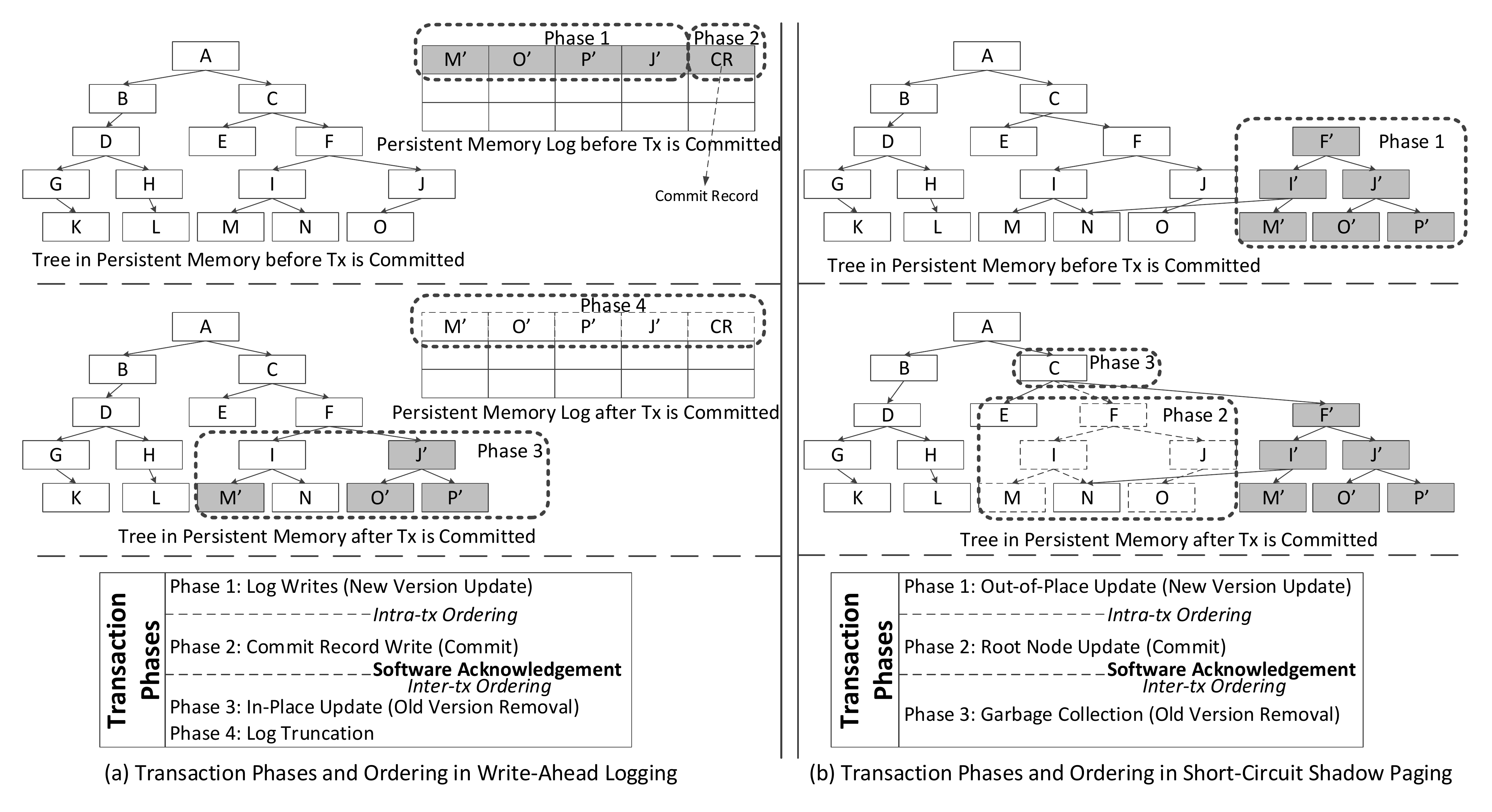}
    \vspace{-0.20in}
    \caption{Transaction Phases and Ordering in (a) Write-Ahead Logging and 
    (b) Short-Circuit Shadow Paging (a variant of Shadow Paging that is used in BPFS~\cite{sosp09bpfs}): This example is illustrated with a tree
      data structure in persistent memory. In the transaction shown in the
      example, page P is inserted and page M, O and J are updated.}
    \label{fig:txrecovery}
\vspace{-0.20in}
\end{figure*}

\textbf{Transaction Recovery.}  Transaction recovery requires
data blocks modified by one transaction to be {\em atomically
  persisted} to storage devices, such that the persistent data blocks
can be used to recover the system to a consistent state {\em after} an
unexpected system crash/failure. To enable this, existing transaction
recovery protocols maintain 1) two copies/versions of each written
data within a transaction, and 2) a strict write order to the storage
device, which enables the atomic switch from the old version of data
to the new version upon transaction commit. We briefly describe
Write-Ahead Logging (WAL)~\cite{tods92aries}, the commonly-used
protocol which we use as our baseline.

\emph{Write-Ahead Logging (WAL)}~\cite{tods92aries} is a
commonly used protocol for transaction recovery.  A transaction commit
occurs in four phases to ensure correct recoverability of data, as
illustrated in the left half of Figure~\ref{fig:txrecovery}. In
Phase 1 (during transaction execution), WAL writes the new version of
each updated data block to a \emph{log area} in persistent memory, while
the old version is kept safe in its home/original location. In Phase 2
(which starts right after the program issues a transaction commit
request), WAL first waits until all the data blocks the transaction
has updated are written into the log. After this, WAL writes a commit
record to the log to keep the transaction status.  At the end of Phase
II, the new-version data and the commit record are persisted
completely and WAL sends an acknowledgment to the program indicating
that the transaction commit is done. In Phase 3, WAL copies the new
version of each updated data block from the log to its home location
to make it visible to accesses from the software (this is called
in-place update of data). Finally, after in-place update completes, in
Phase 4, WAL truncates the log such that the committed transaction is
removed from the log. We call each of these phases an \emph{I/O phase}.

\emph{Shadow paging} also consists of I/O phases, but it is based on 
out-of-place updates. In phase one, the
new-version data are updated in \emph{newly allocated} space. The root node is
written only after the updates of its descendants complete, as shown in
phase two. Like the commit record, the root node serves as the consensus
between normal execution and recovery process. Once the system crashes,
the new-version data are accessible only if the root has been updated.
In phase three, the old-version data can be garbage collected only
after the update of the root. The right half of Figure
\ref{fig:txrecovery} shows a variant of shadow paging, Short-Circuit Shadow
Paging~\cite{sosp09bpfs}, which is used in persistent memory that is
byte addressable and that supports 64-bit atomic writes.

In addition to transaction-based recovery, which we adopt in this paper,
\emph{Soft Updates} is a technique to provide consistency by writing data
blocks that have dependencies in an ordered way~\cite{tocs00soft, atc00softupdates}.
\emph{Soft Updates} is different from transactions in that it does not
keep both the old and new data versions and cannot be recovered to the
old version after system crashes. However, it still has to keep the ordering
of writes and has the ordering overhead as in transaction-based recovery.
More recently, the ThyNVM~\cite{micro15thynvm} consistency model has been
proposed to provide crash consistency to unmodified programs, but it requires
a significant departure from the storage interface to load/store interface to persistent memory.

\textbf{Ordering.}  To achieve atomicity and durability, I/O
phases are performed one by one, in strict order. This is done to
ensure correct recovery in case the system fails/crashes during
transaction commit. Updates to persistent memory across the I/O phases
are performed in a strict order such that one phase cannot be started
before the previous phase is complete. This is called
\emph{persistence ordering}.  Note that this is different from the
ordering of program instructions (loads and stores), which is enforced
by the CPU. Persistence ordering is the ordering of {\em cache
  writebacks to persistent memory} such that the correct ordering of
storage transactions is maintained. As shown in
Figure~\ref{fig:txrecovery},
there are two kinds of persistence ordering in transaction recovery.

\emph{Intra-transaction (Intra-tx) Ordering} refers to the ordering
required within a transaction. Before a transaction commits, WAL needs
to ensure that the new version of each updated data block of the
transaction is completely persisted. Only after that, WAL updates the
commit record. Otherwise, if the commit record is updated before all
data blocks updated by the transaction are persisted, the transaction
recovery process, after a system crash, may incorrectly conclude that
the transaction is committed, violating atomicity and consistency
guarantees. Intra-tx ordering ensures that the new versions of data
are completely and safely written to persistent memory when the commit
record is found during the transaction recovery process.

\emph{Inter-transaction (Inter-tx) Ordering} refers to the ordering
required across transactions. The program needs to wait for the commit
acknowledgment (shown as ``Software Acknowledgment'' in
Figure~\ref{fig:txrecovery}) of a transaction in order to start the next
transaction. Inter-tx ordering ensures that the transaction commit
order is the same as the order specified by the program.

\vspace{-0.1in}
\section{Loose-Ordering Consistency}
\label{sec:design}

Loose-Ordering Consistency (LOC) is designed to mitigate the
performance and endurance degradation caused by strict ordering of writes by
loosening the ordering without compromising consistency in persistent
memory. It aims to reduce both intra-tx and inter-tx ordering
overheads. LOC consists of two techniques:
\begin{enumerate}
\item \emph{Eager Commit}, a commit protocol that eliminates the use of
commit records, thereby removing intra-tx ordering.
\item \emph{Speculative Persistence} that allows writes from different
  transactions to speculatively update persistent memory in any
  order while making them visible to software only in program order,
  thereby relaxing inter-tx ordering.
\end{enumerate}

This section describes both techniques in detail after giving an
overview of the LOC mechanism.

\vspace{-0.1in}
\subsection{Design Overview}
\label{sec:overview}

\begin{figure}[!htb]
    \centering
    \vspace{-0.05in}
    \includegraphics[width=0.76\linewidth]{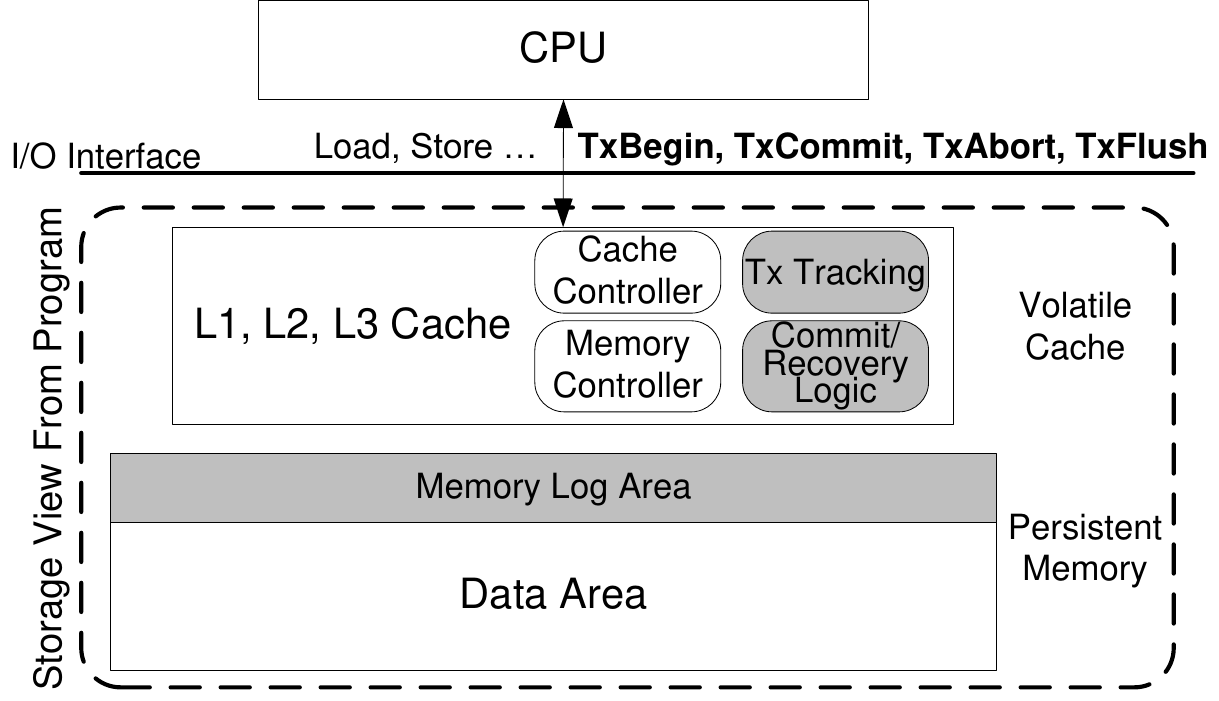}
    \vspace{-0.15in}
    \caption{LOC Design Overview.}
    \vspace{-0.05in}
    \label{fig:overview}
\end{figure}

Figure~\ref{fig:overview} shows an
overview of the LOC design.  CPU issues load and store instructions to
perform memory I/O operations. From the program's point of view, the
volatile CPU cache and the persistent main memory are not
differentiated; all stores to memory within a storage transaction
are deemed to be persistent
memory updates. In order to keep the storage system that resides in
persistent memory consistent, both the I/O interface and the CPU cache
hardware are extended to make sure that the data in the volatile cache is
persisted to persistent memory atomically.

\textbf{Interface.}
The I/O interface, which lies between the CPU core and the CPU cache
hardware, is extended with transactional instructions: TxBegin, TxCommit,
TxAbort and TxFlush. \textit{TxBegin}, \textit{TxCommit} and
\textit{TxAbort} are respectively used to start, commit and abort a
transaction.  \textit{TxFlush} is used to explicitly write data to
persistent memory from the CPU cache. In LOC, durability and atomicity
are decoupled, similarly to the approaches of \cite{sosp13optfs, sosp09bpfs,
cmupdltr11clc}.
\textit{TxCommit} and \textit{TxFlush} are combined to
provide both atomicity and durability in LOC.

\textbf{Components.}
LOC also extends the cache and memory controllers to support transactional
operations and allocates a dedicated memory area for logging.
Figure~\ref{fig:overview} shows the extended components (as highlighted).
\emph{Memory Log Area} is a memory area that is used for keeping
logs and is directly managed by the commit and recovery logic. Memory log
area is \emph{not} visible to the programs. To exploit the parallelism for logging
performance, memory log area spans across different memory banks.
\emph{Transaction~(Tx) Tracking} component has two roles. First, it tracks
all transactional writes. Transactional writes are the writes wrapped between
TxBegin and TxCommit/TxAbort. Each transactional write is tagged
with its TxID.
Second, the Tx Tracking component tracks the status of each transaction
and the commit sequence of running transactions.
\emph{Commit and Recovery Logic} commits or aborts a transaction
and initializes the recovery process if the system crashes. The commit logic directly
manages the memory log area and enforces the ordering between different
transaction phases. During the recovery period after system failures, the
recovery logic reads logs from the memory log area, checks the status of
transactions, and redoes the committed transactions while discarding the uncommitted ones.

\textbf{Operations.}
A transaction writes data in three phases: execution, logging and
checkpointing. In the \textit{execution} phase, data are written to the
CPU cache. In this phase, transactional semantics are passed to the
CPU cache with the extended transactional interface.  Transactional
writes are buffered in the CPU cache until the transaction commits.
When a transaction commits, it enters the \textit{logging} phase, in
which data are persisted to the memory log area (i.e., log write).
Only after all data are completely persisted to the log, a transaction
can enter the \textit{checkpointing} phase, in which data are written
to their home locations in persistent memory (i.e., in-place write).

\begin{figure*}[htb]
\center
\includegraphics[width=0.80\linewidth]{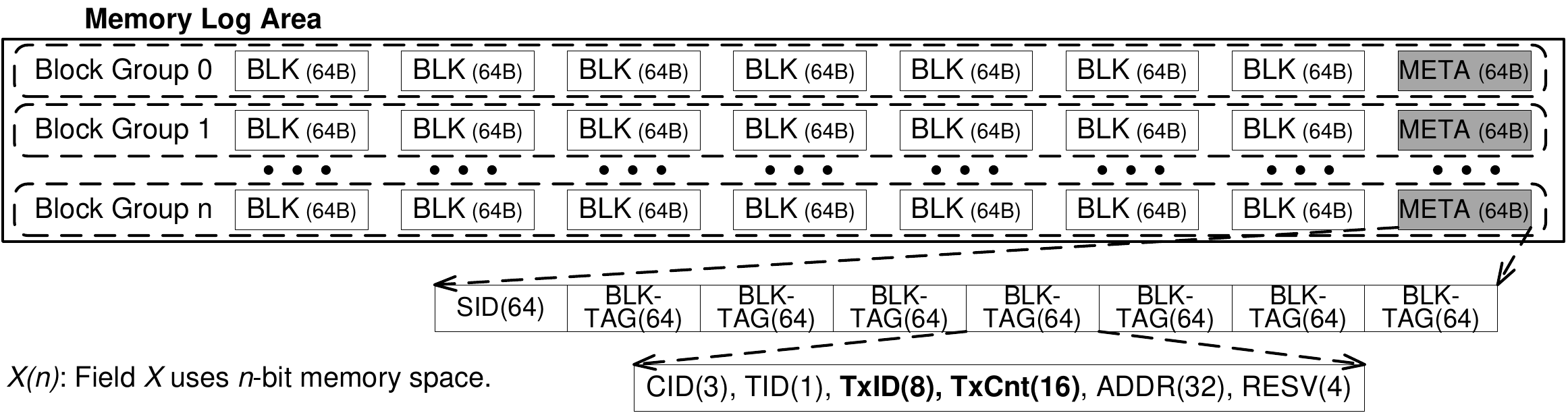}
\vspace{-0.15in}
\caption{Memory Log Organization: Every eight blocks form a block group, where
one is used for metadata and the others for data.}
\label{fig:logformat}
\vspace{-0.20in}
\end{figure*}

In LOC, rather than keeping two copies (log write and in-place write) of
each data block in the CPU cache, LOC only stores a single copy by directly
managing the Memory Log Area.
To improve both performance and endurance of persistent memory, LOC
provides the Eager Commit technique (to remove the intra-tx
ordering at the end of logging phase) and the Speculative Persistence technique (to relax the
inter-tx ordering just after the logging phase).
We describe the two techniques in detail next.

\vspace{-0.15in}
\subsection{Eager Commit}
\label{sec:eager-commit}

The commit protocol in storage transactions is the consensus between
normal execution and recovery on system failure. It is used to
determine when to switch between the \emph{old} and \emph{new} versions of data that
is updated by a transaction. In the commonly used WAL protocol
(described in Section~\ref{sec:background}), a commit
record is used for each transaction to indicate this switch. The
commit protocol in WAL makes sure that (1) the new version of data is
persisted before writing the commit record, and (2) the old version of
the data is overwritten only after the persistent update of the commit
record.  On a system crash/failure, the recovery logic checks the
availability of the commit record for a transaction. If the commit
record exists, the transaction is determined to be committed, and the
new versions of the committed data blocks are copied from the log to
their home locations in persistent memory; otherwise, the transaction
is determined to be not committed (i.e., system might have crashed
before the transaction commit is complete), and the log data
associated with the transaction is discarded.

Unfortunately, it is the commit record itself that introduces the {\em
  intra-tx ordering requirement} (described in
Section~\ref{sec:background}) and therefore degrades
performance heavily in persistent memory. \emph{Eager Commit}
eliminates the use of the commit record and thus removes the intra-tx
ordering. The key idea is to {\em not} wait for the completeness of
log writes and instead eagerly commit a transaction. The completion
check of log writes is delayed until the recovery phase. The removal
of completion check from the critical commit path removes the intra-tx
ordering and thus reduces the commit latency of each transaction. \emph{Eager
Commit} enables a delayed completion check at recovery time using a
\emph{static log organization} with a count-based commit protocol, which we
describe next. This static log organization enables 1) the system to
determine the status of each transaction during recovery without
requiring the use/query of a commit record, 2) enables updates of
different transactions to the log to be interleaved in the log.

\textbf{Log Organization.}  \emph{Eager Commit} organizes the memory
log space in a static manner, as opposed to appending all updated data
blocks and the commit record at the end of the log for a transaction,
as done in WAL. It divides the memory log area into block groups, as
shown in Figure~\ref{fig:logformat}. Each block group consists of
eight data blocks, seven for the log data and one for the metadata
associated with the seven data blocks.
The size of each block is 64 bytes, which can be transmitted to memory
in a single burst~\cite{intelmanual, micro15gsdram}.
In a block group, the eight blocks are issued \emph{serially} by the memory controller, with simple enhancements to the memory controller as in~\cite{micro14firm, micro16delegated}.\footnote{Note that memory controllers are becoming increasingly more intelligent and complex to deal with various scheduling and performance management issues in multi-core and heterogeneous systems (e.g.,~\cite{micro07stfm, isca08parbs, hpca10atlas, micro10tcm, micro11imps, micro11pams, isca12sms, mutlu-memcon13, hpca13mise, iccd14bliss, taco16dash, isca16enhancedmc, micro16continuousrunahead, moscibroda2007memory, kim2012case, subramanian2015application, subramanian2016bliss, lee2008prefetch, ahn2015scalable, ahn2015pim, hsieh2016transparent, hsieh2016accelerating, kim2014bounding, boroumand2016lazypim, chang2016low, jog2016exploiting, david2011memory, seshadri2013rowclone, lee2013tiered, lee2010dram, ipek2008self, moscibroda2008distributed, hpca16chargecache, micro13linearly, pact15decoupled, kim2016bounding, isca15page, hpca15adaptive, hpca14improving, isca12raidr,
hpca17softmc, sigmetrics16understanding, taco16simultaneous, pact15exploiting, isca13orchestrated, hpca16case, 
weed13pm, micro14managing, asplos10fairness, isca11prefetch}).  As such, the additional complexity needed to serially issue eight blocks, as required by our mechanism, has become relatively easy to incorporate, among all the more complex scheduling and performance tasks memory controllers of today deal with.}
This way, writes of the data and
metadata blocks are ordered in each block group, ensuring that all
data blocks are persisted first \emph{before} metadata is persisted, which
guarantees consistency.
Since this ordering is ensured by the memory scheduling mechanism in the memory controller,
it has negligible overhead compared to the ordering in the multi-level CPU cache.
During recovery after a system
crash, the metadata of the block group can be read to determine the
status of the data blocks of the block group.

In a block group, the metadata block stores the sequence ID
(\textit{SID}),
which is the unique number in the memory log area to represent a block
group, and the metadata (\textit{BLK-TAG}) of the other blocks.
\textit{BLK-TAG} records
the CPU core ID (\textit{CID}), the hardware thread ID (\textit{TID})\footnote{The
\textit{CID} and \textit{TID} fields are not used in this paper, but are reserved to support multi-thread and multi-core transactions in future work.},
the transactional
identifier (\textit{TxID}), the transactional counter (\textit{TxCnt}), and the home
location address (\textit{ADDR}) of the data in the block.  The first three IDs
are used to identify the transaction. Therefore, blocks from
different transactions can be written to the log in an interleaved
manner.
For the same reason, log blocks are allocated when the log data is
evicted from the CPU cache to the memory controller. This
ensures that the memory controller buffers do not overflow when
buffering log blocks of the same transaction.
On recovery, each data block is identified as belonging to
some transaction using the three IDs. Afterwards, the commit protocol
uses the pairs $<$$TxID$, $TxCnt$$>$ to determine the transaction
status, as described below.
Finally, LOC retrieves the ADDR of a data
block, which is a pointer that points to the home location in the data area
of a data block, to write the committed data block in the memory log area to its
home location in the data area.

Our current LOC implementation supports 128 running transactions using an 8-bit TxID\footnote{The 256 slots ($[0, 255]$) of TxIDs are used in a circular manner. Only 128 consecutive slots are valid at one time.
For the valid 128 consecutive slots, either $[p, p+127]$ where $0 \le p \le 128$ or $[p, 255] \cup [0, p-129]$ where $128 < p \le 255$, it is easier to compare two TxIDs to find the larger one.},
and allows the largest transaction size of 32K blocks using a 16-bit TxCnt.
The maximum memory space that is currently supported is 256 gigabytes,
due to the 32-bit ADDR (for 64-byte blocks). These limitations can be altered by
changing the bits allocated to the different fields or by using the reserved bits.
In addition, we assume a 64-byte granularity for update operations.
While finer-grained byte addressability reduces internal fragmentation,
it leads to higher metadata overhead. We believe this is a general trade-off,
and the 64-byte granularity, which is the size of a cache block in many modern systems,
is feasible for persistent memory transactions.

\textbf{Commit Protocol.}  In the memory log area, each data
block has associated transactional metadata, $<$$TxID$, $TxCnt$$>$. \emph{Eager
Commit} uses the pair $<$$TxID$, $TxCnt$$>$ to determine the
committed/not-committed status of each transaction.  For each
transaction, the last data block has its associated \textit{TxCnt}
value set to the total number of data blocks in its transaction, and
all the others have \textit{TxCnt} set to zero.  During recovery, the
number of data blocks logged for a transaction (those that have the same
\textit{TxID}) is counted and this count is compared with the non-zero
\textit{TxCnt} stored with one of the data blocks. If the count
matches the non-zero \textit{TxCnt} (indicating that \emph{all} of the
transaction's data blocks are already written to the log),
the transaction is deemed to be
committed and the recovery process copies its updated blocks from the
log to the home locations of the blocks in persistent memory.
Otherwise, the transaction is deemed to be not committed and its
entries in the log are discarded. We borrow this count-based commit protocol
from~\cite{iccd13lighttx}, where it is described in more
detail.

{\em Eager Commit} shares the same philosophy with the \emph{torn bit} technique
in Mnemosyne~\cite{asplos11mnemosyne} to distribute the commit record for
a transaction across the individual log records comprising a transaction
and to avoid the sequential dependency between all the individual logs and
the commit record. In the torn bit technique, one bit is reserved in each
data block to indicate the block's status. Thus, only 63 bits can be stored in each
64-bit data block, and the torn bit technique requires bit shifting.
In contrast, {\em Eager Commit} statically
organizes data blocks into groups and does not need to shift
data in the log records. {\em Eager Commit} is more friendly to data writes,
which are mostly byte-aligned.

In conclusion, {\em Eager Commit} removes the intra-tx ordering by using a
static log organization and a count-based commit protocol that can
enable the determination of transaction status upon recovery \emph{without
requiring a commit record}.

\vspace{-0.10in}
\subsection{Speculative Persistence}
\label{subsec:specpers}

Inter-tx ordering guarantees that the commit sequence of transactions in
the storage system is the same as the {\em commit issue order of
  transactions by the program} (i.e., the order in which transaction
commit commands are issued). To maintain this order, all blocks in one
transaction must be persisted to the memory log before any block of a
later transaction is persisted. To ensure this, a cache conflict in
the CPU cache that causes a block of a later transaction to be evicted
must force the eviction of all data blocks of itself and previous
transactions. Thus, inter-tx ordering not only causes significant
serialization of persistent memory requests but it also results in
inefficient utilization of the CPU cache and higher memory traffic,
thereby degrading system performance and endurance.

\emph{Speculative Persistence} relaxes inter-tx ordering by allowing
blocks from different transactions to be written to the persistent memory
log speculatively, out of the software-specified transaction commit
issue order. However, the written blocks become visible to software
only in the software-specified order. As such, the high-level idea is
somewhat similar to out-of-order execution in
modern processors~\cite{ibm67cpu, patt1985hps, mutlu2003runahead}: persistent
memory writes are completed out-of the program-specified transaction
order (within a window) but they are made visible to software in
program transaction commit order. We call this property
\emph{``out-of-order persistence, in-order commit''}.

With \emph{Speculative Persistence}, a transaction starts persisting its data
blocks without waiting for the completion of the persistence of
previous transactions' data blocks. Instead, there is a
\emph{speculation window}, in which all transactions are persisted
out-of-order. The size of the \emph{speculation window} is called
\emph{speculation degree (SD)}. \textit{Speculation degree} defines the
maximum number of transactions that are allowed to persist log blocks
out-of-order.  As such, the inter-tx ordering is relaxed. Relaxed
inter-tx ordering brings two benefits. First, cache conflict of one
data block does not force eviction of all blocks of its and all
previous transactions.  This improves the cache utilization.  Second,
writes from multiple transactions are coalesced when written back to
memory, which leads to lower memory traffic and improved endurance.

\begin{figure}[htb]
\center
\includegraphics[width=0.99\linewidth]{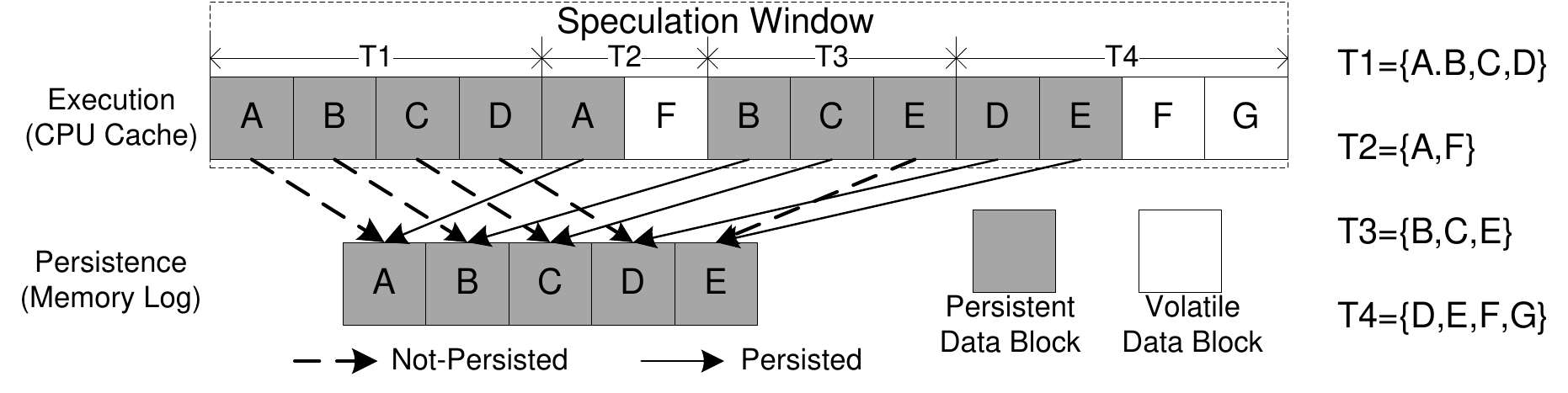}
\vspace{-0.15in}
\caption{Illustration of Speculative Persistence.}
\label{fig:speculativepersistence}
\vspace{-0.15in}
\end{figure}

Figure~\ref{fig:speculativepersistence} illustrates transaction
persistence in a \emph{speculation window} with a \emph{speculation
degree} of four.
Within the \emph{speculation window}, data blocks from the four transactions
can be persisted in any order. For instance, blocks in T3 can be
persisted before blocks in T2. For a data block that has multiple
versions across transactions, only the latest version needs to be
persisted. None of the blocks in T1 (A, B, C, D) need to be written to memory
because their later versions, block A in T2 and blocks B, C, D in T3, will overwrite them.
Therefore, inter-tx ordering is relaxed within the \emph{speculation
window}.

Figure~\ref{fig:speculativepersistence} also illustrates that
\emph{Speculative Persistence} preserves the ``out-of-order persistence,
in-order commit'' property: transaction T1 is reported to be committed
while T3 is not, because T2 has not finished its updates to the
persistent memory log.  To preserve this property, \emph{Speculative
Persistence} has to carefully deal with 1) overlapping writes from
different transactions (to the same block) to ensure that any write to
any block is recoverable, and 2) commit dependencies between
transactions to ensure that transactions are committed in program
order. To enable the former, our mechanism supports multi-versioning
in the CPU cache. To enable the latter, our proposal not only
leverages multi-versioning in the CPU cache but also keeps track
of the committed transaction ID based on the commit issue order of
transactions by the program. We describe both of these next.

\textbf{Multiple Versions in the CPU Cache.}
In \emph{Speculative Persistence}, multiple versions of a data block are
maintained in the volatile CPU cache, similarly to the versioning
cache~\cite{hpca98svc}. Otherwise, if only a
single copy (the latest copy) is kept but the transaction that last
wrote to the block aborts, all previously committed transactions that
were supposed to write to the block would also {\em have to be
  aborted}. With multiple versions of a data block present in the
cache, one version can be removed only when one of its succeeding
versions has been committed (i.e., the software has committed a
later-in-program-order transaction that writes to the data block).
This is because the committed version in a future
transaction that is committed later in program order is guaranteed to
overwrite any previous version regardless of whether the transaction
that is supposed to write to the previous version is aborted or
committed.

There are two issues with keeping multiple versions in the cache: 1)
{\em version overflow}, 2) increased cache pressure. First, version
overflow refers to the case that the associativity of the cache is not
enough to keep all active versions of a cache block (or different
cache blocks) in the corresponding cache set. When this happens, our
design evicts the oldest version to the memory log to make space for
the new version. This eviction can reduce the benefit of \emph{Speculative
Persistence}: probability of merging of writes from different
transactions reduces and an old version of the block may be written to
persistent memory unnecessarily. However, this eviction does \emph{not}
affect the correctness of the commit protocol. Second, since multiple
versions of a block are kept in the cache set, the pressure on the
cache set increases compared to a conventional cache, which may lead
to higher miss rates.
Both of these two cases hurt the efficiency of the CPU cache.\footnote{If only a single copy is kept in the CPU cache (without a versioning
cache), the speculative persistence technique is similar to the \emph{group
commit} technique (or transaction batching)~\cite{hptr88groupcommit}. Transactions in one
speculative window are either all committed or all aborted. With
multiple versions in the CPU cache, the speculative persistence
technique can commit each transaction in the speculative window
separately in a fine-grained manner, which obeys the semantics of
program transactions.
} As a result, memory
traffic increases, accelerating the wear-out process of persistent
memory. Therefore, arbitrarily increasing the speculative degree does
not always bring more benefit and in fact it can do harm by wearing
out memory faster.
Note that, although \emph{Speculative Persistence} keeps multiple cache
versions of a block, only the latest version is persisted to the
memory log when versions do not overflow.
As such, write coalescing is enabled across transactions
within a \emph{speculation window}. In general, memory traffic can be
reduced by using speculative persistence, and memory endurance can be improved.

\textbf{Commit Dependencies Between Transactions.}  Write
coalescing for a block across different transactions causes new
transaction dependencies that need to be resolved carefully at
transaction commit time.  This happens due to two reasons: 1) an
aborted transaction may have overwritten a block in its preceding
transactions, 2) an aborted transaction may have a block that is
overwritten by succeeding transactions that have completed the update
of their logs with the new version. Both reasons are caused
by what we call as \emph{overlapped writes}, i.e., multiple writes 
in which part or all of a data block are written by multiple transactions.
To maintain {\em out-of-order
  persistence, in-order commit} property of \emph{Speculative
Persistence},
we have to deal with two problems when a transaction aborts: 1)
how to rescue the preceding transactions that have overlapped writes
with the aborted transaction?, and 2) how to abort the succeeding
transactions that have completed the write of their logs for an
overlapped write with the aborted transaction?

The two problems are solved by tracking the {\em commit issue order}
of the transactions within each \emph{speculation window} along with
leveraging the multi-versioning support in the CPU cache.  To solve the
first problem, when an abort happens, preceding transactions that have
overlapped writes with the aborted transaction write their versions of
the blocks written by the aborted transaction from the CPU cache to
the persistent memory log.  To solve the second problem, we simply abort
the transactions that come later in the commit issue order than the
aborted transactions.
In the recovery phase, a transaction is
determined to be committed {\em only if} it is checked to be committed
using the count-based commit protocol that checks \textit{TxCnt} (as
described in Section~\ref{sec:eager-commit}) {\em and} its preceding
transactions in its \emph{speculation window} are committed.

{\bf Modifications to the Commit Protocol.} In \emph{Speculative
Persistence}, overlapped writes in the same \emph{speculation window} are
merged, as described above. As a result, a transaction that has
overlapped writes with succeeding transactions does not write the
overlapped data blocks to the persistent memory log. This requires a
modification to the commit protocol we described in
Section~\ref{sec:eager-commit} because, without modification of the
commit protocol, the transaction might be mistaken as a not-committed
transaction as the actual number of data blocks in the memory log area
that are stored for it is different from the non-zero \textit{TxCnt}.  To
differentiate between the two kinds of transactions (those with
overlapped writes and non-overlapped writes), we add a new field of
metadata, \emph{Transaction Dependency Pair}, in the memory log
to represent the dependency between transactions with overlapped
writes. \emph{Transaction Dependency Pair} $<$$T_{a}$, $T_{b}$, $n$$>$ represents that
transaction $T_{a}$ has $n$ overlapped writes with its succeeding
transaction $T_{b}$.  Transaction $T_{a}$ is determined to be
committed if and only if $T_{b}$ is committed and the actual number of
log blocks of $T_{a}$ plus $n$ equals its non-zero \textit{TxCnt}.  As such,
\emph{Transaction Dependency Pairs} help the commit status identification of
transactions with coalesced writes.

\vspace{-0.10in}
\subsection{Recovery from System Failure}

Upon a system crash/failure, LOC scans the memory log area to recover
the system to a consistent state. It first checks the start and end
addresses of the valid logs, and then reads and processes the logs in
the unit of \emph{speculation window}. It processes each
\emph{speculation window}
one by one, in program order. Since strict ordering is required
between \emph{speculation windows}, transactions across
\emph{speculation windows}
have no dependencies. Each \emph{speculation window} can thus be recovered
independently.

First, the \textit{META} block of each data block group is read in
each \emph{speculation window},. LOC counts the number of \textit{BLK-TAG}s
for each $<$$CID$, $TID$, $TxID$$>$; this is the number of logged blocks of
a transaction. If the number matches the non-zero \textit{TxCnt} in
any \textit{BLK-TAG} of the transaction, the transaction is marked as
committed.

Second, {\em Transaction Dependency Pairs} from the memory log area are read.
For each pair $<$$T_{a}$, $T_{b}$, $n$$>$, LOC adds the value $n$ to $T_{a}$
if $T_{b}$ is committed. These pairs are checked in the reverse
sequence, from tail to head.  After this step, transactions that have overlapped writes
are marked as committed using the commit protocol.

Third, the first not-committed transaction is found. All transactions
after it are marked as not-committed. This guarantees the {\em
  in-order commit} property.  After this, LOC finishes the committed
transactions by writing the data blocks of these transactions to the
home locations of the data blocks. Recovery completes after discarding
all data blocks of the not-committed transactions from the log, and
the system returns to a consistent state.


\begin{figure*}[htb]
  \begin{minipage}[b]{0.68\textwidth}
    \centering
    \includegraphics[width=0.9\textwidth]{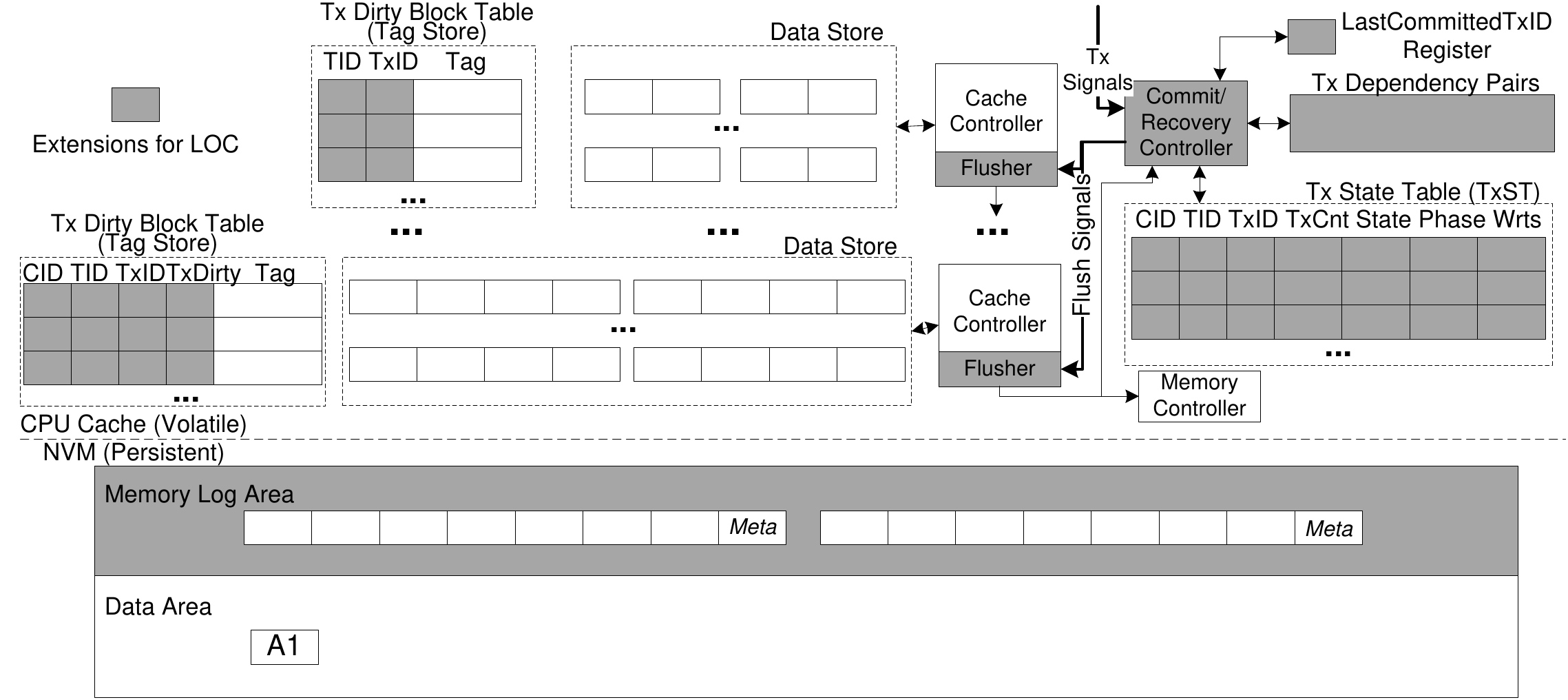}
  \scriptsize{\\(a) LOC Hardware Extensions}
  \end{minipage}%
  \begin{minipage}[b]{0.28\textwidth}
    \centering
\begin{minipage}[b]{\textwidth}
\scriptsize
\begin{tabular}{|c|c|c|c|c|} \hline
(Bits) & L1 & L2 & L3 & TxST \\ \hline
CID & 0 & 0 & 3 & 3 \\
TID & 1 & 1 & 1 & 1  \\
TxID & 8 & 8 & 8 & 8 \\
TxDirty & 0 & 0 & 1 & 0 \\
TxCnt & 0 & 0 & 0 & 16 \\
State  & 0 & 0 & 0 & 2 \\
Phase  & 0 & 0 & 0 & 2 \\
Wrts  & 0 & 0 & 0 & 16 \\ \hline
Total  & 9 & 9 & 12 & 48 \\ \hline
\end{tabular}
\scriptsize{\\ \\ (b) Storage Overhead of Hardware Extensions (extra bits
  per cache block or TxST entry) \\ \\ }
  \vspace{-0.05in}
\end{minipage}
\scriptsize
\begin{minipage}[b]{0.46\textwidth}
    \begin{tabular}{|p{0.08\linewidth}|p{0.45\linewidth}|} \hline
     Bit  & State \\ \hline
      0   & invalid \\
      1   & active \\
      2   & committed \\
      3	  & aborted \\ \hline
    \end{tabular}
\scriptsize{\\ \\ (c) Tx States}
\end{minipage}
\begin{minipage}[b]{0.48\textwidth}
    \begin{tabular}{|p{0.08\linewidth}|p{0.48\linewidth}|} \hline
      Bit & Phase \\ \hline
      0   & log write \\
      1   & in-place \\
		& write \\
      2   & complete \\ \hline
    \end{tabular}
\scriptsize{\\ \\ (d) Tx Phases }
\end{minipage}
  \end{minipage}
\vspace{-0.15in}
\caption{LOC Hardware Extensions and Their Storage Overhead.}
\label{fig:impl}
\vspace{-0.2in}
\end{figure*}

\vspace{-0.15in}
\subsection{LOC vs. Transactional Memory}

As discussed in Section~\ref{sec:background}, Transactional Memory (TM)~\cite{isca93tm,book:tm2} and LOC work respectively on the concurrency control and transaction recovery aspects of transaction mechanisms.  LOC focuses on how to
provide crash consistency in the presence of a transactional interface
(i.e., how to recover transaction states, crash recovery) and does not
deal with how concurrency control is handled. In contrast, TM focuses
on how to provide concurrency control. We currently assume a
transactional interface for LOC (which supports storage transactions),
but do not provide support for TM as our goal is to ensure LOC
works with storage transactions. The combination of TM concurrency
control mechanisms and LOC's transaction recovery mechanisms is left
for future work.

LOC focuses on crash recovery of single-thread transactions, and does not currently support cross-core transactions. In LOC, a unique transaction can be performed only in one core at a given point in time. Concurrency of multi-cores can still be exploited, when there are enough threads from the software performing \emph{different} transactions. Supporting concurrency control is complex due to the need for modifying cache coherence to provide isolation, which is the focus of transactional memory techniques~\cite{isca93tm,book:tm2}, but not LOC. While LOC and transactional memory share similar designs (e.g., a transactional interface and version control mechanisms) and work on different aspects of transactional execution mechanisms, they have potential to be combined to provide full ACID properties (as illustrated in Figure~\ref{fig:tx-pm}). We leave this exploration for future work.

\vspace{-0.15in}
\section{Implementation and Hardware Overhead}
\label{sec:impl}

We now describe the architecture implementation details and discuss
the hardware overhead of LOC.

\vspace{-0.15in}
\subsection{ISA Extensions}
To support storage transactions, We extend the Instruction Set Architecture (ISA) with transactional
instructions: \emph{TxBegin}, \emph{TxCommit} and \emph{TxAbort}.
The ISA extensions are similar to those in hardware transactional memory,
which have been implemented in processors like Intel's
Haswell~\cite{book:intelmanual} and IBM's zEC12~\cite{micro12ibmsystemz}.
Different from transactional memory, ISA extensions in LOC require
persistence in each phase of a transaction.
In LOC, software defines the semantics of storage consistency and specifies the range of a transaction using existing transactional interfaces as in DBMSs, filesystems and other applications: \textit{txbegin, txcommit, txabort}~\cite{tods92aries, osdi08txflash, asplos11mnemosyne, iccd13lighttx}.  A compiler can easily translate these interface commands into the extended ISA instructions, so as to enforce the software's transactional semantics/calls at the hardware level.
Note that LOC allows reordering of non-transactional writes, and does not enforce an ordering between the transactional and non-transactional writes.

In addition, ISA is extended with another instruction, \emph{CheckMaxCommit},
to check the last committed transaction. LOC adds a
\textit{LastCommittedTxID} register in the CPU cache. A committed
transaction updates the register with its ID when all previous
transactions have been committed.
The \emph{CheckMaxCommit} instruction is used for software to query the last
committed transaction.

\noindent\textbf{Limitations.}
LOC does not ensure application consistency after application crashes,
but ensures system consistency for system crashes (e.g., power failure).
This guarantee, i.e., system crash consistency,
is the same as that provided by traditional file systems or database
management systems. In an application crashes, there are two major
types of errors. The first is an incomplete \emph{$<$TxBegin, TxCommit/TxAbort$>$}
pair, i.e., a transaction might not have stopped after being started. This kind of
error can be checked by the compiler and can be reported during
compilation time. The second type is application crash during
execution, i.e., a runtime error. The OS can detect the crashed
process and restart it. Before that, the failed process is killed and
cleaned. Active transactions of threads of the process are
aborted. The transaction (atomicity) semantics ensure the consistent
state of persistent memory.

\vspace{-0.15in}
\subsection{Components}

LOC adds three new components to the system: \textit{Tx Tracking
Component}, \textit{Commit/Recovery Logic}, and \textit{Memory Log Area}.
Figure~\ref{fig:impl} shows where these components reside.

\noindent\textbf{Tx Tracking Component} has two parts: \textit{Tx Dirty
  Block Table} and \textit{Tx State Table}.  \textit{Tx Dirty Block
  Table} tracks the dirty blocks for each transaction at each level of
the CPU cache hierarchy. It is supported with extra bits added to
the tag store of each cache.  As shown in Figure~\ref{fig:impl}(a),
each tag in all cache levels is extended with the hardware thread ID
(\textit{TID}) and the transaction ID (\textit{TxID}).  In the LLC,
two other fields, the CPU core ID (\textit{CID}) and the transaction
dirty flag (\textit{TxDirty}), are also added.  \textit{TxDirty}
indicates whether the block has been written to the persistent memory
log, and the original dirty flag indicates whether the block has been
written to its home location.  Transaction durability is achieved when
log writes are persistent, i.e., \textit{TxDirty} is unset. After that,
home-location writes can be performed using the original dirty flag as
in a conventional CPU cache.  The storage overhead of each cache is
illustrated in Figure~\ref{fig:impl}(b).  Only 9 bits (or 12 bits) are
added for each 64B block in each level of cache (or LLC).

\textit{Tx State Table (TxST)} tracks the status of active
transactions, as shown in Figure~\ref{fig:impl}.  Each \textit{TxST}
entry contains the \textit{CID, TID, TxID, TxCnt, State, Phase and Wrts}
fields.  \textit{State} denotes the transaction state, as shown in
Figure~\ref{fig:impl}(c). State transitions are induced by
transactional commands.  For instance, \textit{TxBegin} changes the
state from invalid to active; \textit{TxCommit} (\textit{TxAbort})
changes the state from active to committed (aborted).  \textit{Phase}
denotes the current status of the write-back of the transaction: the
transaction could be in the \textit{log writing} phase, \textit{in-place
writing} phase updating
home locations or could have completed the entire write-back, as shown
in Figure~\ref{fig:impl}(d).
\textit{Wrts} denotes the number of blocks written back in each phase,
and is used to keep track of the completeness of \textit{log writing} or
\textit{in-place writing} to determine the status of each transaction.  The
storage overhead is shown in Figure~\ref{fig:impl}(b).  Each
\textit{TxST} entry has 48 bits.
For 128 transactions allowed in the
system, the total size of \textit{TxST} is 768 bytes.

\noindent\textbf{Commit/Recovery Logic (CRL)} receives transactional
commands and manages the status of each transaction in the \textit{Tx State
Table}.  CRL stalls new transactions until the current
\textit{speculation window} completes.
For each \emph{speculation window}, CRL tracks different
versions of each data block and adds a 32KB volatile buffer to store
its \emph{Tx Dependency Pairs}.  When a \emph{speculation window} completes, the
buffer is written back to the memory log area. In addition, CRL
maintains a \textit{LastCommittedTxID} register to keep the ID of the
last committed transaction, which indicates to the software that all transactions
with smaller IDs are committed.

\noindent\textbf{Memory Log Area} is a contiguous physical memory
space to log the writes from transactions. At the beginning of the
memory log area, there is a log head, which records the start and end
addresses of the valid data logs.  The main body of the memory log area
consists of the log data block groups (as shown in
Figure~\ref{fig:logformat}) and the metadata of
\textit{Tx Dependency Pairs} (introduced in Section~\ref{subsec:specpers}).
This area is invisible to the software, and is allocated by the memory controller.
To overcome the endurance problem of NVM, allocation of this area is moved around
in the physical memory periodically (similar to the notion of Start-Gap Wear Leveling~\cite{micro09startgap}). Note that other techniques can also be employed to ensure that the
memory log area or the memory does not wear out due to log updates.
In our evaluations, 32MB memory is allocated for the \textit{Memory Log Area} as this
was empirically found to be enough for the supported 128 transactions,
but this space can be dynamically expanded.

\vspace{-0.15in}
\section{Evaluation}
\label{sec:eval}

In this section, we first compare LOC with previous transaction
protocols. Then, we analyze the benefits of {\em Eager
  Commit} and {\em Speculative Persistence}. Finally,
we study sensitivity to transaction size and memory latency.

\vspace{-0.15in}
\subsection{Experimental Setup}

We evaluate different transaction protocols using a full-system
simulator, GEM5~\cite{gem5}. GEM5 is configured using the syscall
emulation (SE) mode.  Benchmarks can directly run on the full system
simulator without modification or recompilation. In the evaluation,
GEM5 uses the {\em Timing Simple CPU} mode and the {\em Ruby memory
  system}. The CPU is 1 GHz, and the CPU cache and memory have the
parameters shown in Table~\ref{table:simulator}.
We revise both the cache and memory controllers in GEM5 to simulate
LOC, as shown in Figure~\ref{fig:impl}.  We
faithfully model all overheads associated with LOC. In our evaluation,
the {\em Speculation Degree} of LOC is set to 16 by default.

\begin{table}[htb]
\vspace{-0.2in}
\caption{\label{table:simulator}Simulator Configuration. }
\vspace{-0.15in}
\centering
  \begin{tabular}{c l}
    \hline
    L1 Cache & 32KB, 2-way associative, 64B block size, \\
		& LRU replacement, block access latency = 1 cycle \\
    L2 Cache & 256KB, 8-way associative, 64B block size \\
		& LRU replacement, block access latency = 8 cycles \\
    LLC & 1MB, 16-way associative, 64B block size \\
		& LRU replacement, block access latency = 21 cycles \\
    Memory & 8 banks, memory access latency = 168 cycles\\
   \hline
  \end{tabular}
\vspace{-0.1in}
\end{table}

{\bf Workloads.} Table~\ref{table:workloads} lists the workloads we
evaluate.  B+ tree is a widely used data structure in both file
systems and database management systems. We implement a B+ tree, in
which each 4KB node contains 200 key(8B)-value(4B) pairs. Each
transaction consists of multiple key-value insert or delete
operations. Similarly, we use the hash table, red-black tree and random
array swap data structures, also used in
literature~\cite{asplos11nvheaps}.  Our graph processing workload
inserts and deletes edges in a large graph~\cite{boost}. We also
evaluate a database workload~\cite{leveldb-bench} on SQLite
3.7.17~\cite{sqlite}.
In Table~\ref{table:workloads}, we provide the average size of the persistence set
(P-Set), which is the number of data blocks per transaction that have to be persisted to ensure transaction
durability, in each workload.

\begin{table}[htb]
\vspace{-0.15in}
\caption{\label{table:workloads}Workloads.}
\vspace{-0.25in}
\centering
\begin{tabular}[t] {l l c}
\hline
\multirow{2}{*}{Workloads} & \multirow{2}{*}{Description} & P-Set \\
 & & (blks/tx)\\
\hline
\hline
B+ Tree & Insert/delete nodes in a B+ tree & 89.60 \\
Hash~\cite{asplos11nvheaps} & Insert/delete entries in a hash table & 10.92\\
RBTree~\cite{asplos11nvheaps} & Insert/delete nodes in a red-black tree & 33.26\\
SPS~\cite{asplos11nvheaps} & Random swaps of array entries & 1.53\\
SDG~\cite{boost} & Insert/delete edges in a large graph & 52.85 \\
SQLite~\cite{leveldb-bench} & Database benchmark on SQLite & 309.34 \\
\hline
\end{tabular}
\vspace{-0.2in}
\end{table}

\vspace{-0.05in}
\subsection{Overall Performance}
\label{subsec:overall}

We measure and compare both transaction throughput and memory write traffic of five different
transaction protocols: S-WAL, H-WAL, LOC-WAL, Kiln and LOC-Kiln. S-WAL
is a software WAL (Write-Ahead Logging) protocol that manages logging and ordering in
software, as shown in Figure~\ref{fig:txrecovery}(a).
H-WAL is a hardware WAL protocol that
manages logging in hardware.  Different from S-WAL, which writes two
copies respectively for log and in-place writes in the CPU cache,
H-WAL keeps only a single copy in the CPU cache and lets the hardware
manage the log and in-place writes. H-WAL does not change the ordering
behavior of S-WAL. LOC-WAL is our proposed protocol in this paper.
Kiln is a recent protocol that uses a non-volatile last-level cache (LLC) to
reduce the persistence overhead~\cite{micro13kiln}. It keeps the new
and old version of data blocks respectively in the LLC and persistent memory,
thereby eliminating the need for performing double copies in persistent memory.
Kiln also needs to preserve the order with which transactional writes are evicted from the LLC only after the transaction commits. Otherwise, the transactional writes overwrite the old-version data in persistent memory, which cannot be recovered after a transaction failure.
Since Kiln's optimization is orthogonal
to our LOC mechanism, we combine the two and also evaluate this
combined version, called LOC-Kiln. LOC-Kiln achieves the best of both
Kiln and LOC by only flushing L1 and L2 caches (as in Kiln) and
performing loose ordering (as in LOC).

\begin{figure}[htb]
\vspace{-0.2in}
\center
\includegraphics[width=0.86\linewidth]{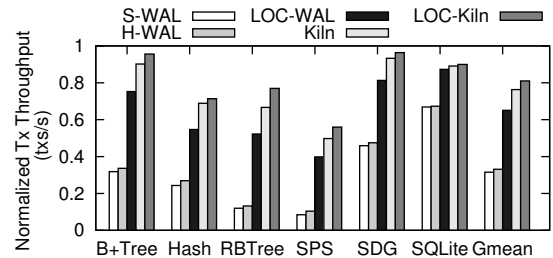}
\vspace{-0.2in}
\caption{Performance Comparison of Consistency Protocols.}
\vspace{-0.1in}
\label{fig:overall-thruput}
\end{figure}

\textbf{Transaction Throughput.}
Figure~\ref{fig:overall-thruput} shows the normalized transaction throughput of
the five protocols.  The results are normalized to the transaction
throughput of the baseline, which runs benchmarks without any
transaction support and thus without the associated overheads of
transactions. Note that this baseline does \emph{not} provide consistency
upon system failure and is intended to show the overhead of providing
such consistency via different mechanisms. Transaction throughput is
calculated by dividing the total number of committed transactions with the total
runtime of each benchmark.  We make two key observations.

(1) LOC significantly improves the performance of WAL, including
both S-WAL and H-WAL. Normalized transaction throughput increases from
0.316 in S-WAL and 0.331 in H-WAL to 0.651 in LOC-WAL. In other words,
LOC-WAL reduces ordering overhead from 68.4\% in S-WAL and 66.9\% in
H-WAL to 34.9\%.  This is because S-WAL manages logging in
software. The log writes and home-location writes have different
memory addresses, and thus are independently cached in the CPU
cache. Keeping two copies in the CPU cache hurts cache efficiency, which
H-WAL removes, but this does not greatly reduce the overhead of WAL.
LOC greatly reduces the overhead of WAL by removing intra-tx ordering
using {\em Eager Commit} and loosening inter-tx ordering using {\em
  Speculative Persistence}. The loosened ordering improves cache
efficiency and increases the probability of write coalescing in the CPU
cache.

(2) LOC and Kiln can be combined favorably. Doing so
improves normalized transaction throughput to 0.811 on average, i.e.,
the ordering overhead is reduced to 18.9\%.  Kiln shortens the
persistence path by employing a non-volatile last-level cache.  LOC
mitigates performance degradation via a complementary technique, i.e.,
loosening the persistence ordering overhead that still exists in a non-volatile
cache.

\begin{figure}[htb]
\vspace{-0.15in}
\center
\includegraphics[width=0.82\linewidth]{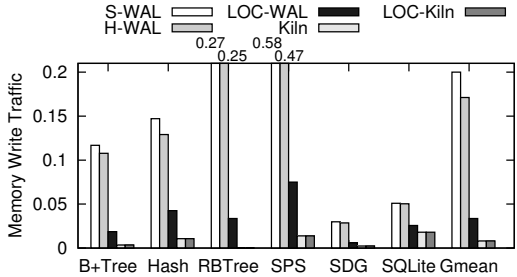}
\vspace{-0.15in}
\caption{Endurance Comparison of Consistency Protocols.}
\vspace{-0.1in}
\label{fig:overall-wrsz}
\end{figure}

\textbf{Endurance.}
Figure~\ref{fig:overall-wrsz} shows the memory write traffic of
the five protocols. The memory write traffic in the figure is calculated by dividing the
total size of memory writes by the total size of program writes.
Lower memory write traffic indicates better cache efficiency of the CPU
cache and better endurance of persistent memory.
As shown in the figure, the memory write traffic in different workloads
is dramatically reduced from 0.200 in H-WAL and 0.171 in S-WAL to 0.034
in LOC-WAL on average.
The benefits of LOC-WAL come from the write coalescing of overlapped transactional writes
across transactions in the speculative persistence technique,
which will be further evaluated in Section~\ref{subsec:eval-sp}.
Kiln uses a non-volatile last level cache, and the memory write traffic is
only 0.008. LOC-Kiln has even lower memory write traffic, which is
0.06\% lower than that in Kiln.

We conclude that LOC effectively mitigates performance degradation and
reduces write traffic (and thus improves endurance of persistent memory)
due to persistence ordering by relaxing both intra- and
inter-transaction ordering.

\vspace{-0.15in}
\subsection{Effect of the Eager Commit Protocol}

\begin{figure}[htb]
\center
\vspace{-0.15in}
\includegraphics[width=0.76\linewidth]{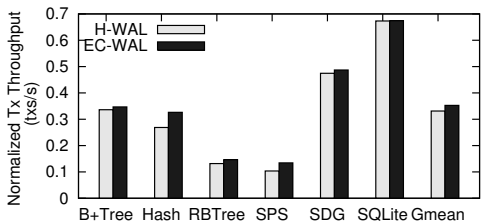}
\vspace{-0.15in}
\caption{Effect of Eager Commit on Transaction Throughput.}
\label{fig:eagercommit}
\vspace{-0.2in}
\end{figure}

We compare the transaction throughput of H-WAL and EC-WAL. EC-WAL is
the LOC mechanism for WAL with only {\em Eager Commit} but without
{\em Speculative Persistence}.  Figure~\ref{fig:eagercommit} plots the
normalized transaction throughput of the two techniques. EC-WAL
outperforms H-WAL by 6.4\% on average. This is because it removes the
completeness check in {\em Eager Commit} from the critical
path of transaction commit.  The elimination of intra-tx ordering
leads to fewer cache flushes and improves cache efficiency.

\vspace{-0.05in}
\subsection{Effect of Speculative Persistence}
\label{subsec:eval-sp}

To evaluate both performance and endurance gains from {\em Speculative Persistence}, we
vary the \emph{speculation degree (SD)} from 1 to 32 (SD was set to 16 in
previous evaluations).

\begin{figure}[htb]
\center
\vspace{-0.2in}
\includegraphics[width=0.85\linewidth]{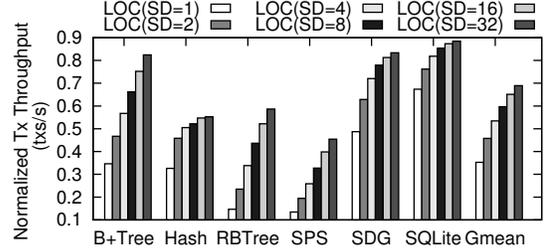}
\vspace{-0.15in}
\caption{Effect of Speculative Persistence on Transaction Throughput.}
\label{fig:loc-perf}
\vspace{-0.15in}
\end{figure}

Figure~\ref{fig:loc-perf} shows the normalized
transaction throughput of LOC-WAL with different SD values. On
average, the normalized transaction throughput of LOC-WAL increases
from 0.353 to 0.689 with 95.5\% improvement, going from SD=1 to SD=32. This
benefit comes from two aspects of \emph{Speculative Persistence}.  First,
\emph{Speculative Persistence} allows out-of-order persistence of different
transactions. A cache block without a cache conflict is \emph{not} forced to
be written back to persistent memory within a \emph{speculation window} (as
explained in Section~\ref{subsec:specpers}), thereby reducing memory traffic and improving
cache efficiency. Second, \emph{Speculative Persistence} enables write
coalescing \emph{across transactions} within the \emph{speculation window}, thereby reducing
memory traffic.  Both of these effects increase as the \emph{speculation
degree} increases, leading to larger performance benefits with larger
\emph{speculation degrees}.

\begin{figure}[htb]
\center
\vspace{-0.15in}
\includegraphics[width=0.85\linewidth]{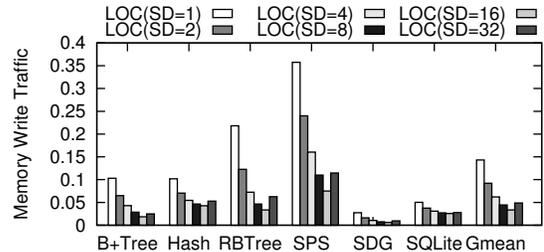}
\vspace{-0.15in}
\caption{Effect of Speculative Persistence on Write Traffic.}
\label{fig:loc-wrsz}
\vspace{-0.1in}
\end{figure}

Figure~\ref{fig:loc-wrsz} shows the memory write traffic of LOC-WAL with
different speculation degrees.
In each workload, the memory write traffic is reduced instantly
when the speculation degree (SD) is increased from 1 to 16.
This shows that the probability of write coalescing gets higher
when more transactions are allowed to be persisted out-of-order.
An exception to this trend is observed when SD is 32. When SD increases from 16 to 32,
LOC-WAL shows a slight increase in memory write traffic. The reason is
that more concurrently executed transactions incur a higher probability of
cache evictions in the CPU cache. Generally, memory write traffic
reduces with a higher speculation degree.

In conclusion, LOC generally gains more benefits when the speculation degree is
higher, i.e., when more transactions are allowed to be persisted
concurrently.

\vspace{-0.15in}
\subsection{Impact of Transaction Size}

To study the impact of transaction size on performance penalty from
ordering, we also measure the transaction throughput of LOC-WAL using
B+ Tree with different transaction sizes. Transaction size is changed by
updating a different number of tree nodes in each transaction, varying
from 2 to 32.

\begin{figure}[htb]
\center
\vspace{-0.10in}
\includegraphics[width=0.76\linewidth]{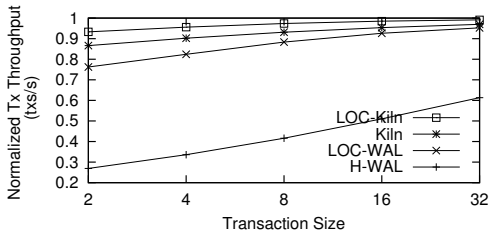}
\vspace{-0.15in}
\caption{Impact of Transaction Size.}
\label{fig:impact-txsz}
\vspace{-0.1in}
\end{figure}

Figure~\ref{fig:impact-txsz} shows the normalized transaction throughput
of LOC-WAL under workloads with different transaction sizes.
Results in each transaction size setting are normalized to the baseline case,
which runs benchmarks without transaction support.
Two conclusions are in order.
First, the LOC mechanism is more beneficial to workloads with smaller transaction
sizes than those with larger ones. When the transaction size is small,
amortized ordering overhead for each operation is high. Therefore,
ordering overhead is more significant in workloads with smaller transaction
sizes. Thus, relaxed ordering in the LOC mechanism has a larger effect.
Second, performance gains from non-volatile last-level cache (LLC) are less
significant than those from the LOC mechanism when transaction size is large.
When the transaction size becomes larger, not all data blocks can be buffered
in the LLC. As Kiln requires that all new-version data are
persistently buffered in the LLC, Kiln has to fall back to WAL mode for memory
logging for large transactions. Thus, Kiln provides less benefit for
workloads with larger transaction sizes. When the transaction size increases to
32, both LOC-WAL and LOC-Kiln achieve performance comparable to baseline without
any transaction overhead.

In conclusion, the LOC mechanism is especially beneficial to workloads with small transaction sizes, yet it provides performance benefits across the board, for workloads with many different transaction sizes.

\vspace{-0.15in}
\subsection{Sensitivity to Memory Latency}

\begin{figure}[htb]
\center
\vspace{-0.15in}
\includegraphics[width=0.99\linewidth]{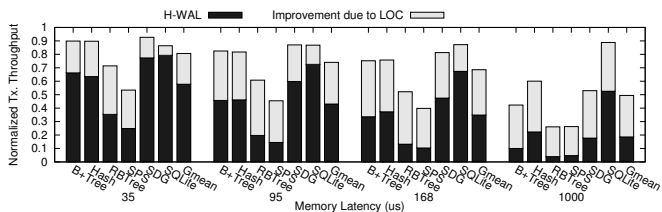}
\vspace{-0.2in}
\caption{Sensitivity to Memory Latency.}
\label{fig:impact-memlatency}
\vspace{-0.15in}
\end{figure}

We evaluate LOC performance with different memory latencies to
approximate the effect of different types of non-volatile memories.
We vary the memory latency between 35, 95, 168 and 1000 nanoseconds (our
default evaluations so far were with a 168-nanosecond latency). We
measure the transaction throughput of both H-WAL and LOC at each
latency.  Figure~\ref{fig:impact-memlatency} shows the performance
improvement of LOC over H-WAL at different memory latencies. In the
figure, the black part of each stacked bar shows the normalized
transaction throughput of H-WAL, and the grey part shows the
performance improvement of LOC over H-WAL. Two major observations are
in order.  First, performance of H-WAL reduces as memory latency
increases.  This shows that higher memory latency in NVMs leads to
higher persistence ordering overheads.  Second, LOC's performance
improvement increases as memory latency increases.  This is because
LOC is able to reduce the persistence overhead, which increases with
memory latency.  We conclude that persistence ordering overhead is
becoming a more serious issue with higher-latency NVMs, which LOC can
effectively mitigate.

\vspace{-0.1in}
\section{Related Work}
\label{sec:related}

\textbf{Mitigating the Ordering Overhead.}
As explained in Section~\ref{sec:intro}, in order to preserve
persistence ordering from the CPU cache to persistent memory, software
combines cache flush (e.g., \textit{clflush}) and memory fence (e.g.,
\textit{mfence}) instructions to force the ordering of cache
writebacks~\cite{sc11scmfs, asplos11nvheaps, asplos11mnemosyne,
  fast11cdds}. The average overhead of a clflush and mfence combined
together is reported to be 250ns~\cite{asplos11mnemosyne}, which makes
this approach costly, given that persistent memory access times are
expected to be on the order of
tens to hundreds of nanoseconds~\cite{isca09pcmlee,isca09pcmqureshi,
  ispass13sttram}.\footnote{Recent research argues that these
  commands, which are used for cache coherence, do not correctly flush
  cache data to persistent memory, and thus proposes to ensure
  ordering in CPU hardware~\cite{sosp09bpfs, eurosys14pmfs}.}  The two
instructions flush dirty data blocks from the CPU cache to persistent
memory and wait for the completion of all memory writes, and thus incur
high overhead in persistent memory~\cite{sosp09bpfs,
  asplos11mnemosyne, micro13kiln, isca14memorypersistency}.

Several works attempt to mitigate the ordering overhead in persistent
memory with hardware support.
These can be classified into two approaches, as follows.

(1) {\em Making the CPU cache non-volatile}:
This approach aims to reduce the time gap between volatility and
persistence by employing a non-volatile cache. Kiln~\cite{micro13kiln}
uses a non-volatile last-level cache (NV-LLC), so that the path of
data persistence becomes shorter and the overhead of the required ordering
is smaller. Kiln also uses the NV-LLC as the log to eliminate the need
to perform multiple writes in main memory.
Whole-system persistence~\cite{asplos12wholesystempersistence} takes this approach
to the extreme by making all levels of the CPU cache non-volatile.
Similarly, JUSTDO~\cite{asplos16justdo} makes all levels of the CPU cache
non-volatile and optimizes the log organization.
The approach we develop in this paper, LOC, is complementary to the
approach of employing NV caches.

(2) {\em Allowing asynchronous commit of transactions}:
This approach allows the execution of a later transaction without
waiting for the persistence of previous transactions. To ensure
consistency, the program queries the hardware for the persistence
status of transactions.  BPFS~\cite{sosp09bpfs} and
CLC~\cite{cmupdltr11clc} use versions of this approach. They inform
the CPU cache hardware of the ordering points within the program via
the \textit{epoch} command, and let the hardware keep the ordering
asynchronously without waiting for data persistence within each epoch.
In addition to the cache hierarchy, enhancements have been made to
the memory controller to support the \textit{epoch}
semantics~\cite{micro16delegated, asplos17whisper}.
The \textit{epoch} semantic is also extended to multicore CPUs~\cite{micro15multicore}.
Strand persistency~\cite{isca14memorypersistency} enables the
reordering of the commit sequence of transactions for better
concurrency (instead of requiring a strict order in which transactions
have to be committed).
Mechanisms for isolation and concurrency have also been supported
by enhancing the memory controller in various ways~\cite{hpca16backendcontroller, cf15dp2, micro14firm}.
A recent work, ThyNVM~\cite{micro15thynvm}, introduces checkpointing to the DRAM+NVM hybrid
persistent memory to overlap application execution and hardware
checkpointing, to provide transparent crash consistency.

The \emph{asynchronous commit} approaches change only the execution order
(i.e., the order of CPU cache writes) but \emph{not} the persistence order (i.e.,
the order of persistent memory writes) as specified by the
program.\footnote{Even with strand
persistency~\cite{isca14memorypersistency}, the persistence order is
fixed once transaction concurrency has been specified in the
program.}
In contrast, our proposal, LOC, allows the
reordering of persistent memory writes of different transactions in a
finer-grained manner and can be combined with the asynchronous commit
approaches.

BPPM~\cite{msst15bppm, tos15bppm} reduces the persistence
overhead by blurring the volatility-persistency boundary. This
work relaxes the requirement of data persistence. In contrast, LOC
does not require the relaxation of data persistence
requirements. Instead, it reduces the consistency overhead by
speculatively relaxing the ordering requirement using a hardware-based
approach. Note that LOC can be combined with the method of \cite{msst15bppm, tos15bppm} for
even higher benefits.

\noindent\textbf{Eager and Speculative Techniques.}
Recent studies propose new commit protocols to reduce the overhead caused by strict
ordering in traditional write-ahead logging or shadow paging mechanisms.
In file systems research, NoFS~\cite{fast12nofs} is proposed to use backpointers
for pointer-based consistency. It is suitable for tree-like structures,
e.g., file system metadata, but not for general transactions.
OptFS~\cite{sosp13optfs} allows the reordering of log records and the commit record
by storing the checksums of all log records in the commit record.
The checksums can be used to determine the transaction status in the recovery phase, 
similar to the checksum usage in ext4~\cite{ext4}.

In flash-based storage, the no-overwrite property of flash memory requires updates
to be performed in a copy-on-write manner, which naturally simplifies transaction support.
New commit protocols are proposed to eliminate the writes of logs and commit records
by leveraging either the log-structured FTL~\cite{hpca11atomicwrite} or the Out-of-Band
(OOB) area~\cite{iccd13lighttx, fast13ofss, osdi08txflash}.
In contrast to flash-based SSDs, non-volatile memory is byte addressable. NVM neither has an OOB area,
nor does it need to update data out-of-place. Therefore, these techniques cannot be easily or
efficiently applied to persistent memory.

In NVM-based secondary storage, MARS~\cite{sosp13mars} proposes a new interface, editable
atomic writes (EAW), in a PCIe SSD to optimize WAL.
Since the PCIe SSD has the commit and recovery logic implemented inside the SSD
controller, it is not suitable for persistent memory that is directly attached
to the processor through memory interface.
Different from such approaches that apply to storage, our proposed LOC mechanism is designed
for persistent memory.	

In persistent memory, Mnemosyne~\cite{asplos11mnemosyne} introduces the torn-bit
technique, which uses one bit in each block to indicate whether the
block has been written to. The use of the torn-bit removes the usage
of commit records.
A recent work~\cite{asplos16commitlock} proposes to defer transaction commit operations
until after locks are released rather than performing such operations while holding the
locks, to reduce persist dependencies for concurrently running transactions.
While the two works~\cite{asplos11mnemosyne, asplos16commitlock} require changes to the software, LOC is a hardware approach
leveraging the memory organization.

Speculative techniques have been used in many architectural or system
designs, including memory systems to reduce the overhead of memory
fences for concurrent programs~\cite{isca09invisifence, hpca98svc}.
These works deal with different problems from LOC. While these techniques are proposed
for program or instruction concurrency (the ordering of program execution),
speculative persistence in LOC is designed for the ordering of IO persistence.

\vspace{-0.10in}
\section{Conclusion}
\label{sec:concl}

Persistent memory provides disk-like data persistence at DRAM-like
latencies, but requires memory writes to be written to persistent
memory in a strict program order to maintain storage consistency.
Enforcing such strict persistence ordering requires flushing dirty
blocks from all levels of the volatile CPU caches and waiting for
their completion at the persistent memory, which dramatically degrades
system performance. Forced flush operations also incur high memory write
traffic, which hurts the endurance of persistent memory.
To mitigate this performance and endurance overhead, we
introduced \emph{Loose-Ordering Consistency (LOC)}, which relaxes the
persistence ordering requirement without compromising storage
consistency. LOC's two key mechanisms, {\em Eager Commit} and {\em
  Speculative Persistence}, in combination, relax write ordering
requirements both within a transaction and across multiple
transactions. Our evaluations show that LOC can greatly improve system
performance and endurance by reducing the ordering overhead across a wide variety of
workloads. LOC also combines favorably with non-volatile CPU caches,
providing performance benefits on top of systems that employ
non-volatile last-level caches. We conclude that LOC can provide
a high-performance consistency substrate for future persistent memory
systems.

\vspace{-0.1in}
\ifCLASSOPTIONcompsoc
  \section*{Acknowledgments}
\else
  \section*{Acknowledgment}
\fi

This work is supported by the National Natural Science Foundation of
China (61232003, 61433008), the Beijing Municipal Science and Technology
Commission of China (D151100000815003), Intel Science and Technology Center
for Cloud Computing, US National Science Foundation (CNS 1320531, CCF
1212962), and China Postdoctoral Science Foundation (2016T90094, 2015M580098).
This paper is an extended and revised version of~\cite{iccd14loc}.
Jiwu Shu and Onur Mutlu are the corresponding authors.

\ifCLASSOPTIONcaptionsoff
  \newpage
\fi

\vspace{-0.1in}



\bibliographystyle{IEEEtranS}
\bibliography{paper-simple}

\begin{thebibliography}{100}
\providecommand{\url}[1]{#1}
\csname url@samestyle\endcsname
\providecommand{\newblock}{\relax}
\providecommand{\bibinfo}[2]{#2}
\providecommand{\BIBentrySTDinterwordspacing}{\spaceskip=0pt\relax}
\providecommand{\BIBentryALTinterwordstretchfactor}{4}
\providecommand{\BIBentryALTinterwordspacing}{\spaceskip=\fontdimen2\font plus
\BIBentryALTinterwordstretchfactor\fontdimen3\font minus
  \fontdimen4\font\relax}
\providecommand{\BIBforeignlanguage}[2]{{%
\expandafter\ifx\csname l@#1\endcsname\relax
\typeout{** WARNING: IEEEtranS.bst: No hyphenation pattern has been}%
\typeout{** loaded for the language `#1'. Using the pattern for}%
\typeout{** the default language instead.}%
\else
\language=\csname l@#1\endcsname
\fi
#2}}
\providecommand{\BIBdecl}{\relax}
\BIBdecl

\bibitem{ext4}
``Ext4,'' https://ext4.wiki.kernel.org/.

\bibitem{leveldb-bench}
``{LevelDB} benchmarks,''
  \url{http://leveldb.googlecode.com/svn/trunk/doc/benchmark.html}.

\bibitem{sqlite}
``Sqlite,'' \url{http://www.sqlite.org}.

\bibitem{boost}
``Graph: adjacency list,'' \url{http://www.boost.org/doc/libs/1_52_0}, 2013.

\bibitem{ahn2015scalable}
J.~Ahn \emph{et~al.}, ``A scalable processing-in-memory accelerator for
  parallel graph processing,'' in \emph{ISCA}, 2015.

\bibitem{ahn2015pim}
J.~Ahn \emph{et~al.}, ``{PIM}-enabled instructions: A low-overhead,
  locality-aware processing-in-memory architecture,'' in \emph{ISCA}, 2015.

\bibitem{isca12sms}
R.~Ausavarungnirun \emph{et~al.}, ``Staged memory scheduling: Achieving high
  performance and scalability in heterogeneous systems,'' in \emph{ISCA}, 2012.

\bibitem{pact15exploiting}
R.~Ausavarungnirun \emph{et~al.}, ``Exploiting inter-warp heterogeneity to
  improve {GPGPU} performance,'' in \emph{PACT}, 2015.

\bibitem{gem5}
N.~Binkert \emph{et~al.}, ``The {Gem5} simulator,'' \emph{SIGARCH Computer
  Architecture News}, 2011.

\bibitem{isca09invisifence}
C.~Blundell \emph{et~al.}, ``Invisifence: Performance-transparent memory
  ordering in conventional multiprocessors,'' in \emph{ISCA}, 2009.

\bibitem{boroumand2016lazypim}
A.~Boroumand \emph{et~al.}, ``{LazyPIM}: An efficient cache coherence mechanism
  for processing-in-memory,'' \emph{IEEE CAL}, 2016.

\bibitem{sigmetrics16understanding}
K.~K. Chang \emph{et~al.}, ``Understanding latency variation in modern {DRAM}
  chips: Experimental characterization, analysis, and optimization,'' in
  \emph{SIGMETRICS}, 2016.

\bibitem{chang2016low}
K.~K. Chang \emph{et~al.}, ``Low-cost inter-linked subarrays {(LISA)}: Enabling
  fast inter-subarray data movement in {DRAM},'' in \emph{HPCA}, 2016.

\bibitem{hpca14improving}
K.~K.-W. Chang \emph{et~al.}, ``Improving {DRAM} performance by parallelizing
  refreshes with accesses,'' in \emph{HPCA}, 2014.

\bibitem{inflow16onur}
H.~Chauhan \emph{et~al.}, ``{NVMOVE}: Helping programmers move to byte-based
  persistence,'' in \emph{INFLOW}, 2016.

\bibitem{sosp13optfs}
V.~Chidambaram \emph{et~al.}, ``Optimistic crash consistency,'' in \emph{SOSP},
  2013.

\bibitem{fast12nofs}
V.~Chidambaram \emph{et~al.}, ``Consistency without ordering,'' in \emph{FAST},
  2012.

\bibitem{sosp13mars}
J.~Coburn \emph{et~al.}, ``{From ARIES to MARS}: Transaction support for
  next-generation solid-state drives,'' in \emph{SOSP}, 2013.

\bibitem{asplos11nvheaps}
J.~Coburn \emph{et~al.}, ``{NV-Heaps}: making persistent objects fast and safe
  with next-generation, non-volatile memories,'' in \emph{ASPLOS}, 2011.

\bibitem{sosp09bpfs}
J.~Condit \emph{et~al.}, ``Better {I/O} through byte-addressable, persistent
  memory,'' in \emph{SOSP}, 2009.

\bibitem{david2011memory}
H.~David \emph{et~al.}, ``Memory power management via dynamic voltage/frequency
  scaling,'' in \emph{ICAC}, 2011.

\bibitem{dac09pdram}
G.~Dhiman \emph{et~al.}, ``{PDRAM}: a hybrid {PRAM} and {DRAM} main memory
  system,'' in \emph{DAC}, 2009.

\bibitem{hpca16backendcontroller}
K.~Doshi \emph{et~al.}, ``Atomic persistence for scm with a non-intrusive
  backend controller,'' in \emph{HPCA}, 2016.

\bibitem{eurosys14pmfs}
S.~R. Dulloor \emph{et~al.}, ``System software for persistent memory,'' in
  \emph{EuroSys}, 2014.

\bibitem{asplos10fairness}
E.~Ebrahimi \emph{et~al.}, ``Fairness via source throttling: a configurable and
  high-performance fairness substrate for multi-core memory systems,'' in
  \emph{ASPLOS}, 2010.

\bibitem{isca11prefetch}
E.~Ebrahimi \emph{et~al.}, ``Prefetch-aware shared-resource management for
  multi-core systems,'' in \emph{ISCA}, 2011.

\bibitem{micro11pams}
E.~Ebrahimi \emph{et~al.}, ``Parallel application memory scheduling,'' in
  \emph{MICRO}, 2011.

\bibitem{tocs00soft}
G.~R. Ganger \emph{et~al.}, ``Soft updates: a solution to the metadata update
  problem in file systems,'' \emph{TOCS}, 2000.

\bibitem{hpca98svc}
S.~Gopal \emph{et~al.}, ``Speculative versioning cache,'' in \emph{HPCA}, 1998.

\bibitem{acm81shadowpaging}
J.~Gray \emph{et~al.}, ``The recovery manager of the {System R} database
  manager,'' \emph{ACM Computing Surveys}, 1981.

\bibitem{book:tm2}
T.~Harris \emph{et~al.}, ``Transactional memory,'' \emph{Synthesis Lectures on
  Computer Architecture}, 2010.

\bibitem{isca16enhancedmc}
M.~Hashemi \emph{et~al.}, ``Accelerating dependent cache misses with an
  enhanced memory controller,'' in \emph{ISCA}, 2016.

\bibitem{micro16continuousrunahead}
M.~Hashemi \emph{et~al.}, ``Continuous runahead: Transparent hardware
  acceleration for memory intensive workloads,'' in \emph{MICRO}, 2016.

\bibitem{hpca16chargecache}
H.~Hassan \emph{et~al.}, ``Chargecache: Reducing {DRAM} latency by exploiting
  row access locality,'' in \emph{HPCA}, 2016.

\bibitem{hpca17softmc}
H.~Hassan \emph{et~al.}, ``{SoftMC}: A flexible and practical open-source
  infrastructure for enabling experimental {DRAM} studies.''\hskip 1em plus
  0.5em minus 0.4em\relax HPCA, 2017.

\bibitem{hptr88groupcommit}
P.~Helland \emph{et~al.}, ``Group commit timers and high volume transaction
  systems,'' Hewlett-Packard Laboratory, Tech. Rep., 1989.

\bibitem{isca93tm}
M.~Herlihy and J.~E.~B. Moss, ``Transactional memory: Architectural support for
  lock-free data structures,'' in \emph{ISCA}, 1993.

\bibitem{hsieh2016transparent}
K.~Hsieh \emph{et~al.}, ``Transparent offloading and mapping {(TOM)}: Enabling
  programmer-transparent near-data processing in gpu systems,'' in \emph{ISCA},
  2016.

\bibitem{hsieh2016accelerating}
K.~Hsieh \emph{et~al.}, ``Accelerating pointer chasing in {3D}-stacked memory:
  Challenges, mechanisms, evaluation,'' in \emph{ICCD}, 2016.

\bibitem{book:intelmanual}
Intel, ``Intel architecture instruction set extensions programming reference,
  319433-015,'' 2013.

\bibitem{intelmanual}
Intel, ``Intel{\textregistered} 64 and {IA-32} architectures software
  developer’s manual,'' 2014.

\bibitem{ipek2008self}
E.~Ipek \emph{et~al.}, ``Self-optimizing memory controllers: A reinforcement
  learning approach,'' in \emph{ISCA}, 2008.

\bibitem{asplos16justdo}
J.~Izraelevitz \emph{et~al.}, ``Failure-atomic persistent memory updates via
  justdo logging,'' in \emph{ASPLOS}, 2016.

\bibitem{micro12ibmsystemz}
C.~Jacobi \emph{et~al.}, ``Transactional memory architecture and implementation
  for ibm system z,'' in \emph{MICRO}, 2012.

\bibitem{isca13orchestrated}
A.~Jog \emph{et~al.}, ``Orchestrated scheduling and prefetching for {GPGPUs},''
  in \emph{ISCA}, 2013.

\bibitem{jog2016exploiting}
A.~Jog \emph{et~al.}, ``Exploiting core criticality for enhanced {GPU}
  performance.'' in \emph{SIGMETRICS}, 2016.

\bibitem{micro15multicore}
A.~Joshi \emph{et~al.}, ``Efficient persist barriers for multicores,'' in
  \emph{MICRO}, 2015.

\bibitem{ics13memorage}
J.-Y. Jung and S.~Cho, ``Memorage: Emerging persistent {RAM} based malleable
  main memory and storage architecture,'' in \emph{ICS}, 2013.

\bibitem{micro14managing}
O.~Kayiran \emph{et~al.}, ``Managing {GPU} concurrency in heterogeneous
  architectures,'' in \emph{MICRO}, 2014.

\bibitem{kim2014bounding}
H.~Kim \emph{et~al.}, ``Bounding memory interference delay in {COTS}-based
  multi-core systems,'' in \emph{RTAS}, 2014.

\bibitem{kim2016bounding}
H.~Kim \emph{et~al.}, ``Bounding and reducing memory interference in
  {COTS}-based multi-core systems,'' \emph{Journal of Real-Time Systems}, 2016.

\bibitem{hpca10atlas}
Y.~Kim \emph{et~al.}, ``{ATLAS}: A scalable and high-performance scheduling
  algorithm for multiple memory controllers,'' in \emph{HPCA}, 2010.

\bibitem{micro10tcm}
Y.~Kim \emph{et~al.}, ``Thread cluster memory scheduling: Exploiting
  differences in memory access behavior,'' in \emph{MICRO}, 2010.

\bibitem{kim2012case}
Y.~Kim \emph{et~al.}, ``A case for exploiting subarray-level parallelism
  {(SALP)} in {DRAM},'' in \emph{ISCA}, 2012.

\bibitem{asplos16commitlock}
A.~Kolli \emph{et~al.}, ``High-performance transactions for persistent
  memories,'' in \emph{ASPLOS}, 2016.

\bibitem{micro16delegated}
A.~Kolli \emph{et~al.}, ``Delegated persist ordering,'' in \emph{MICRO}, 2016.

\bibitem{ispass13sttram}
E.~Kultursay \emph{et~al.}, ``Evaluating {STT-RAM} as an energy-efficient main
  memory alternative,'' in \emph{ISPASS}, 2013.

\bibitem{isca09pcmlee}
B.~C. Lee \emph{et~al.}, ``Architecting phase change memory as a scalable
  {DRAM} alternative,'' in \emph{ISCA}, 2009.

\bibitem{cacm10pcm}
B.~C. Lee \emph{et~al.}, ``Phase change memory architecture and the quest for
  scalability,'' \emph{Communications of the ACM}, 2010.

\bibitem{ieeemicro10pcm}
B.~C. Lee \emph{et~al.}, ``Phase-change technology and the future of main
  memory,'' \emph{IEEE Micro}, 2010.

\bibitem{lee2008prefetch}
C.~J. Lee \emph{et~al.}, ``Prefetch-aware {DRAM} controllers,'' in
  \emph{MICRO}, 2008.

\bibitem{lee2010dram}
C.~J. Lee \emph{et~al.}, ``{DRAM}-aware last-level cache writeback: Reducing
  write-caused interference in memory systems,'' \emph{UT Austin Tech Report},
  2010.

\bibitem{micro09membank}
C.~J. Lee \emph{et~al.}, ``Improving memory bank-level parallelism in the
  presence of prefetching,'' in \emph{MICRO}, 2009.

\bibitem{taco16simultaneous}
D.~Lee \emph{et~al.}, ``Simultaneous multi-layer access: Improving {3D}-stacked
  memory bandwidth at low cost,'' \emph{TACO}, 2016.

\bibitem{hpca15adaptive}
D.~Lee \emph{et~al.}, ``Adaptive-latency {DRAM}: Optimizing {DRAM} timing for
  the common-case,'' in \emph{HPCA}, 2015.

\bibitem{lee2013tiered}
D.~Lee \emph{et~al.}, ``Tiered-latency {DRAM}: A low latency and low cost
  {DRAM} architecture,'' in \emph{HPCA}, 2013.

\bibitem{pact15decoupled}
D.~Lee \emph{et~al.}, ``Decoupled direct memory access: Isolating {CPU and IO}
  traffic by leveraging a dual-data-port {DRAM},'' in \emph{PACT}, 2015.

\bibitem{isca12raidr}
J.~Liu \emph{et~al.}, ``{RAIDR}: Retention-aware intelligent {DRAM} refresh,''
  in \emph{ISCA}, 2012.

\bibitem{usenix17octopus}
Y.~Lu \emph{et~al.}, ``Octopus: an {RDMA}-enabled distributed persistent memory
  file system,'' in \emph{USENIX ATC}, 2017.

\bibitem{iccd13lighttx}
Y.~Lu \emph{et~al.}, ``{LightTx}: A lightweight transactional design in
  flash-based {SSDs} to support flexible transactions,'' in \emph{ICCD}, 2013.

\bibitem{tctxssd}
Y.~Lu \emph{et~al.}, ``High-performance and lightweight transaction support in
  flash-based {SSDs},'' \emph{IEEE Transactions on Computers}, 2015.

\bibitem{tcdifftx}
Y.~Lu \emph{et~al.}, ``Supporting system consistency with differential
  transactions in flash-based {SSDs},'' \emph{IEEE Transactions on Computers},
  2016.

\bibitem{msst15bppm}
Y.~Lu \emph{et~al.}, ``Blurred persistence in transactional persistent
  memory,'' in \emph{MSST}, 2015.

\bibitem{tos15bppm}
Y.~Lu \emph{et~al.}, ``Blurred persistence: Efficient transactions in
  persistent memory,'' \emph{TOS}, 2016.

\bibitem{iccd14loc}
Y.~Lu \emph{et~al.}, ``Loose-ordering consistency for persistent memory,'' in
  \emph{ICCD}, 2014.

\bibitem{fast14reconfs}
Y.~Lu \emph{et~al.}, ``{ReconFS}: A reconstructable file system on flash
  storage,'' in \emph{FAST}, 2014.

\bibitem{fast13ofss}
Y.~Lu \emph{et~al.}, ``Extending the lifetime of flash-based storage through
  reducing write amplification from file systems,'' in \emph{FAST}, 2013.

\bibitem{nvmsa14txcache}
Y.~Lu \emph{et~al.}, ``{TxCache}: Transactional cache using byte-addressable
  non-volatile memories in {SSDs},'' in \emph{NVMSA}, 2014.

\bibitem{meza2012enabling}
J.~Meza \emph{et~al.}, ``Enabling efficient and scalable hybrid memories using
  fine-granularity {DRAM} cache management,'' \emph{IEEE CAL}, 2012.

\bibitem{iccd12meza}
J.~Meza \emph{et~al.}, ``A case for small row buffers in non-volatile main
  memories,'' in \emph{ICCD}, 2012.

\bibitem{weed13pm}
J.~Meza \emph{et~al.}, ``A case for efficient hardware/software cooperative
  management of storage and memory,'' in \emph{WEED}, 2013.

\bibitem{tods92aries}
C.~Mohan \emph{et~al.}, ``{ARIES}: a transaction recovery method supporting
  fine-granularity locking and partial rollbacks using write-ahead logging,''
  \emph{TODS}, 1992.

\bibitem{cmupdltr11clc}
I.~Moraru \emph{et~al.}, ``Persistent, protected and cached: Building blocks
  for main memory data stores,'' PDL, CMU, Tech. Rep., 2011.

\bibitem{moscibroda2007memory}
T.~Moscibroda and O.~Mutlu, ``Memory performance attacks: Denial of memory
  service in multi-core systems,'' in \emph{USENIX Security}, 2007.

\bibitem{moscibroda2008distributed}
T.~Moscibroda and O.~Mutlu, ``Distributed order scheduling and its application
  to multi-core {DRAM} controllers,'' in \emph{PODC}, 2008.

\bibitem{micro11imps}
S.~P. Muralidhara \emph{et~al.}, ``Reducing memory interference in multicore
  systems via application-aware memory channel partitioning,'' in \emph{MICRO},
  2011.

\bibitem{mutlu-memcon13}
O.~Mutlu, ``Memory scaling: A systems architecture perspective,'' in
  \emph{MEMCON}, 2013.

\bibitem{micro07stfm}
O.~Mutlu and T.~Moscibroda, ``Stall-time fair memory access scheduling for chip
  multiprocessors,'' in \emph{MICRO}, 2007.

\bibitem{isca08parbs}
O.~Mutlu and T.~Moscibroda, ``Parallelism-aware batch scheduling: Enhancing
  both performance and fairness of shared {DRAM} systems,'' in \emph{ISCA},
  2008.

\bibitem{mutlu2003runahead}
O.~Mutlu \emph{et~al.}, ``Runahead execution: An alternative to very large
  instruction windows for out-of-order processors,'' in \emph{HPCA}, 2003.

\bibitem{superfri14research}
O.~Mutlu and L.~Subramanian, ``Research problems and opportunities in memory
  systems,'' \emph{SUPERFRI}, 2014.

\bibitem{asplos17whisper}
S.~Nalli \emph{et~al.}, ``An analysis of persistent memory use with whisper,''
  in \emph{ASPLOS}, 2017.

\bibitem{asplos12wholesystempersistence}
D.~Narayanan and O.~Hodson, ``Whole-system persistence,'' in \emph{ASPLOS},
  2012.

\bibitem{eurosys16hinfs}
J.~Ou \emph{et~al.}, ``A high performance file system for non-volatile main
  memory,'' in \emph{EuroSys}, 2016.

\bibitem{hpca11atomicwrite}
X.~Ouyang \emph{et~al.}, ``Beyond block {I/O}: Rethinking traditional storage
  primitives,'' in \emph{HPCA}, 2011.

\bibitem{patt1985hps}
Y.~N. Patt \emph{et~al.}, ``{HPS}, a new microarchitecture: rationale and
  introduction,'' in \emph{MICRO}, 1985.

\bibitem{hpca16case}
G.~Pekhimenko \emph{et~al.}, ``A case for toggle-aware compression for {GPU}
  systems,'' in \emph{HPCA}, 2016.

\bibitem{micro13linearly}
G.~Pekhimenko \emph{et~al.}, ``Linearly compressed pages: a low-complexity,
  low-latency main memory compression framework,'' in \emph{MICRO}, 2013.

\bibitem{isca14memorypersistency}
S.~Pelley \emph{et~al.}, ``Memory persistency,'' in \emph{ISCA}, 2014.

\bibitem{osdi08txflash}
V.~Prabhakaran \emph{et~al.}, ``Transactional flash,'' in \emph{OSDI}, 2008.

\bibitem{bookpcm}
M.~K. Qureshi \emph{et~al.}, ``Phase change memory: From devices to systems,''
  \emph{Synthesis Lectures on Computer Architecture}, 2011.

\bibitem{micro09startgap}
M.~K. Qureshi \emph{et~al.}, ``Enhancing lifetime and security of {PCM}-based
  main memory with start-gap wear leveling,'' in \emph{MICRO}, 2009.

\bibitem{isca09pcmqureshi}
M.~K. Qureshi \emph{et~al.}, ``Scalable high performance main memory system
  using phase-change memory technology,'' in \emph{ISCA}, 2009.

\bibitem{book-database}
R.~Ramakrishnan and J.~Gehrke, \emph{Database management systems}.\hskip 1em
  plus 0.5em minus 0.4em\relax Osborne/McGraw-Hill, 2000.

\bibitem{micro15thynvm}
J.~Ren \emph{et~al.}, ``{ThyNVM}: Enabling software-transparent crash
  consistency in persistent memory systems,'' in \emph{MICRO}, 2015.

\bibitem{atc00softupdates}
M.~I. Seltzer \emph{et~al.}, ``Journaling versus soft updates: Asynchronous
  meta-data protection in file systems.'' in \emph{USENIX ATC}, 2000.

\bibitem{isca14dbi}
V.~Seshadri \emph{et~al.}, ``The dirty-block index,'' in \emph{ISCA}, 2014.

\bibitem{seshadri2013rowclone}
V.~Seshadri \emph{et~al.}, ``{RowClone}: fast and energy-efficient in-{DRAM}
  bulk data copy and initialization,'' in \emph{MICRO}, 2013.

\bibitem{micro15gsdram}
V.~Seshadri \emph{et~al.}, ``Gather-scatter {DRAM}: In-{DRAM} address
  translation to improve the spatial locality of non-unit strided accesses,''
  in \emph{MICRO}, 2015.

\bibitem{isca15page}
V.~Seshadri \emph{et~al.}, ``Page overlays: An enhanced virtual memory
  framework to enable fine-grained memory management,'' in \emph{ISCA}, 2015.

\bibitem{inflow16sttram}
M.~Shihab \emph{et~al.}, ``Couture: Tailoring {STT-MRAM} for persistent main
  memory,'' in \emph{INFLOW}, 2016.

\bibitem{iccd14bliss}
L.~Subramanian \emph{et~al.}, ``The blacklisting memory scheduler: Achieving
  high performance and fairness at low cost,'' in \emph{ICCD}, 2014.

\bibitem{subramanian2016bliss}
L.~Subramanian \emph{et~al.}, ``{BLISS}: Balancing performance, fairness and
  complexity in memory access scheduling,'' \emph{TPDS}, 2016.

\bibitem{subramanian2015application}
L.~Subramanian \emph{et~al.}, ``The application slowdown model: Quantifying and
  controlling the impact of inter-application interference at shared caches and
  main memory,'' in \emph{MICRO}, 2015.

\bibitem{hpca13mise}
L.~Subramanian \emph{et~al.}, ``{MISE}: Providing performance predictability
  and improving fairness in shared main memory systems,'' in \emph{HPCA}, 2013.

\bibitem{cf15dp2}
L.~Sun \emph{et~al.}, ``{DP2}: Reducing transaction overhead with differential
  and dual persistency in persistent memory,'' in \emph{CF}, 2015.

\bibitem{ibm67cpu}
R.~M. Tomasulo, ``An efficient algorithm for exploiting multiple arithmetic
  units,'' \emph{IBM J. Res. Dev.}, 1967.

\bibitem{linuxexpo98journaling}
S.~C. Tweedie, ``Journaling the {Linux} ext2fs filesystem,'' in \emph{The
  Fourth Annual Linux Expo}, 1998.

\bibitem{taco16dash}
H.~Usui \emph{et~al.}, ``{DASH}: Deadline-aware high-performance memory
  scheduler for heterogeneous systems with hardware accelerators,''
  \emph{TACO}, 2016.

\bibitem{fast11cdds}
S.~Venkataraman \emph{et~al.}, ``Consistent and durable data structures for
  non-volatile byte-addressable memory,'' in \emph{FAST}, 2011.

\bibitem{asplos11mnemosyne}
H.~Volos \emph{et~al.}, ``Mnemosyne: lightweight persistent memory,'' in
  \emph{ASPLOS}, 2011.

\bibitem{sc11scmfs}
X.~Wu and A.~L.~N. Reddy, ``{SCMFS}: a file system for storage class memory,''
  in \emph{SC}, 2011.

\bibitem{iccd12hybrid}
H.~Yoon \emph{et~al.}, ``Row buffer locality aware caching policies for hybrid
  memories,'' in \emph{ICCD}, 2012.

\bibitem{yoon2015efficient}
H.~Yoon \emph{et~al.}, ``Efficient data mapping and buffering techniques for
  multilevel cell phase-change memories,'' \emph{TACO}, 2015.

\bibitem{micro13kiln}
J.~Zhao \emph{et~al.}, ``Kiln: Closing the performance gap between systems with
  and without persistence support,'' in \emph{MICRO}, 2013.

\bibitem{micro14firm}
J.~Zhao \emph{et~al.}, ``{FIRM}: Fair and high-performance memory control for
  persistent memory systems,'' in \emph{MICRO}, 2014.

\bibitem{isca09pcmzhou}
P.~Zhou \emph{et~al.}, ``A durable and energy efficient main memory using phase
  change memory technology,'' in \emph{ISCA}, 2009.

\end{thebibliography}
%
%
%

%

\vspace{-0.30in}
\begin{IEEEbiography}[{\includegraphics[width=1in,height=1.25in,clip,keepaspectratio]{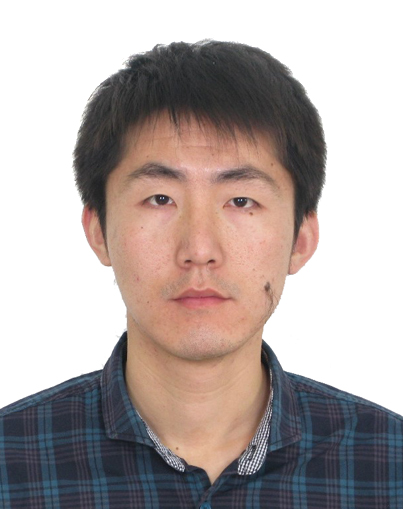}}]{Youyou
Lu}
is a postdoctoral researcher at the Department of Computer Science
and Technology, Tsinghua University. His current research interests include non-volatile memories and file systems.
He obtained his B.S. degree in Computer Science from Nanjing University
in 2009, and his Ph.D degree in Computer Science from Tsinghua
University in 2015.
He is a Member of the IEEE.
\end{IEEEbiography}

\vspace{-0.4in}
\begin{IEEEbiography}[{\includegraphics[width=1in,height=1.25in,clip,keepaspectratio]{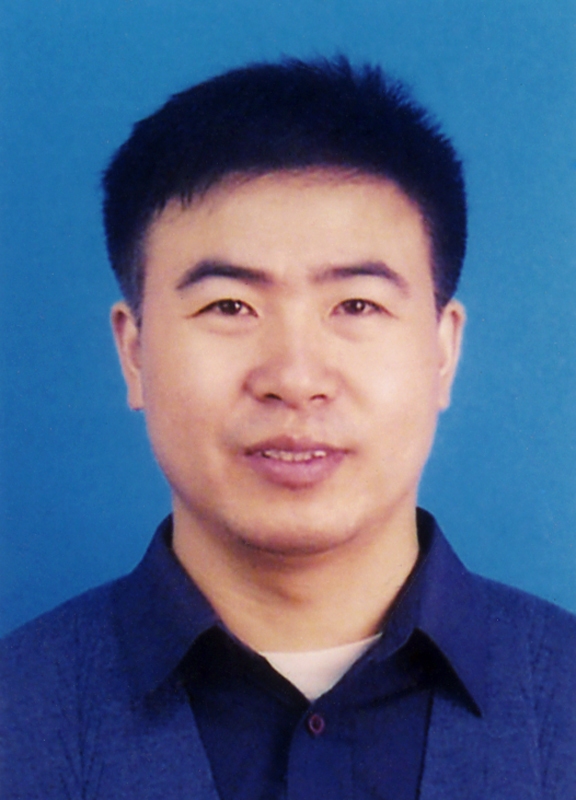}}]{Jiwu
Shu}
is a professor in the Department of Computer Science and Technology
at Tsinghua University.
His current research interests include storage security and reliability,
non-volatile memory based storage systems, and parallel and distributed
computing.
He obtained his Ph.D degree in Computer Science from Nanjing University
in 1998, and finished his postdoctoral research at Tsinghua
University
in 2000. Since then, he has been teaching at Tsinghua University.
He is a Senior Member of the IEEE.
\end{IEEEbiography}

\vspace{-0.4in}
\begin{IEEEbiography}[{\includegraphics[width=1in,height=1.25in,clip,keepaspectratio]{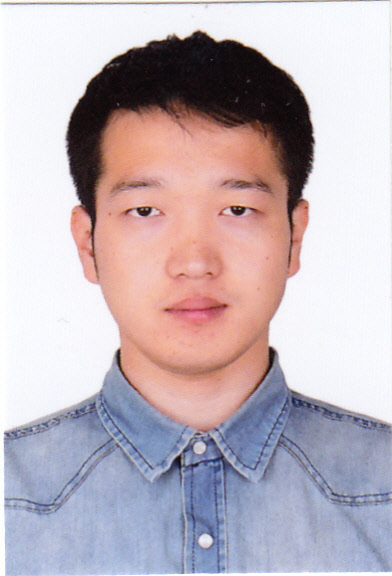}}]{Long Sun}
received his Master degree in the Department of Computer Science and Technology
at Tsinghua University in 2015, and his B.S. degree in Computer Science from Zhejiang University in
2012.
His research interests include non-volatile memories.
\end{IEEEbiography}

\vspace{-0.4in}
\begin{IEEEbiography}[{\includegraphics[width=1in,height=1.25in,clip,keepaspectratio]{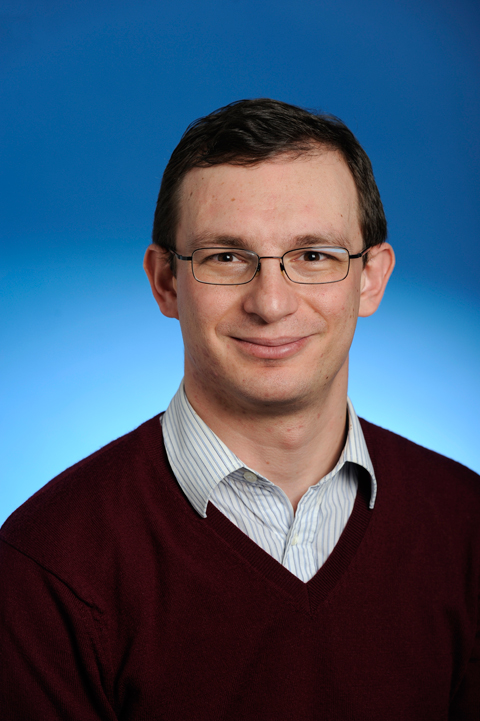}}]{Onur
Mutlu}
is a Full Professor of Computer Science at ETH Zurich. He is also a faculty member at Carnegie Mellon University, where he previously held the William D. and Nancy W. Strecker Early Career Professorship. His current broader research interests are in computer architecture, systems, and bioinformatics. He is especially interested in interactions across domains and between applications, system software, compilers, and microarchitecture, with a major current focus on memory and storage systems. He obtained his Ph.D. and M.S. in ECE from the University of Texas at Austin and B.S. degrees in Computer Engineering and Psychology from the University of Michigan, Ann Arbor. His industrial experience spans starting the Computer Architecture Group at Microsoft Research (2006-2009), and various product and research positions at Intel Corporation, Advanced Micro Devices, Google, and VMware.
He received the inaugural IEEE Computer Society Young Computer Architect Award, the inaugural Intel Early Career Faculty Award, faculty partnership awards from various companies, and a healthy number of best paper or “Top Pick” paper recognitions at various computer systems and architecture venues. His computer architecture course lectures and materials are freely available on YouTube, and his research group makes software artifacts freely available online. For more information, please see his webpage at https://people.inf.ethz.ch/omutlu.
\end{IEEEbiography}




\end{document}